\documentclass[journal]{IEEEtran}
\usepackage{setspace}

\usepackage{amsmath, amsthm, amssymb}
\usepackage{latexsym}
\usepackage{amsfonts}
\usepackage{amscd}
\usepackage{caption}
\usepackage{subcaption}
\usepackage{times}
\usepackage{graphics}
\usepackage{graphicx}
\usepackage{epstopdf}
\usepackage{epsfig}
\graphicspath{{figures/}}
\usepackage{color}
\usepackage{cite}
\usepackage{enumerate}
\usepackage{array}
\usepackage{multirow}
\usepackage{slashbox}

\begin{document}

%
\title{Analysis and Control  of Beliefs in Social Networks}

\author{Tian~Wang,
        Hamid~Krim,~\IEEEmembership{Fellow,~IEEE,}
        Yannis~Viniotis
        }

\maketitle

\begin{abstract}
In this paper, we investigate the problem of how beliefs diffuse among members of social networks. We propose an \textit{information flow model} (IFM) of belief that captures how interactions among members affect the diffusion and eventual convergence of a belief. The IFM model includes a generalized Markov Graph (GMG) model as a social network model, which reveals that the diffusion of beliefs depends heavily on two characteristics of the social network characteristics, namely degree centralities and clustering coefficients. We apply the IFM to both converged belief estimation and belief control strategy optimization. The model is compared with an IFM including the Barab\'{a}ási-Albert model, and is evaluated via experiments with published real social network data. 
\end{abstract}


\IEEEpeerreviewmaketitle

\section{Introduction}
\label{intro}

\IEEEPARstart{A} social network, as an abstract model of a social environment, consists of a set of nodes, which could be a set of individuals or a set of groups of individuals, and a set of relationships with specific characteristics among these nodes. There are numerous types of existing social networks in our daily lives; examples include on-line social networks such as Facebook or LinkedIn, email networks and alumni networks. Possibilities for forming new ones abound; for example, social networks among patients with a rare disease. Compared with non-social types of networks (e.g., a sensor network), social networks exhibit certain specific characteristics in quantities such as degree distributions, clustering coefficient distributions, etc.  \cite{C1}, \cite{C10}, \cite{C12}, which allow one to further analyse and control them.

Within a social network, each node has a certain belief representing its current status. The belief of a node may be influenced by other nodes connected to it, and could be changed in time. The belief may be, for example, an opinion regarding the quality of a restaurant, or the preference to attend a school \cite{D3}. In a different example, the belief may be the opinion of a patient regarding the curability of their illness. A node can propagate its belief to other connected nodes via the network links by performing certain activities, such as dining in a certain restaurant or attending a particular school, or simply by informing the other nodes. Such exposure and exchange of beliefs is actually a flow of information in a social network.

The beliefs in a social network may have value  for members of the network or even outsiders. For example, knowing members' opinions about a certain product  can help a product manufacturer better predict the market and form merchandising strategies. In a clinical study, a doctor may be interested in researching ways to influence patients' behaviour by ``facilitating'' interactions among patients. Consequently, prediction or even  control of the beliefs in a social network  can be  an important and interesting problem. As such, it requires a mathematical model to simulate and analyse the flow of  beliefs in a social network.

We call this model an \textit{Information Flow Model} (IFM). Briefly speaking, IFM includes two main parts: a description of how beliefs are updated in time, and a description of the network structure. The first part describes how information of a member's belief flows in the network and influences the other members. 
The second one describes the paths inside the network over which the   beliefs can be transmitted. 
Details about both parts are given in Section~\ref{IFM}.


To date, research in IFMs has been  mainly conducted on social learning \cite{D1}, \cite{D2}, \cite{D3}, \cite{X8}, which focuses on the first part. The effects of the second part have not received much attention, primarily due to the complexity of social network structure. Early studies on IFM had   assumed   acyclic networks \cite{D3}, \cite{D4},\cite{D6}. More recent works adapt simple descriptions of social networks, including some limited connectivity properties \cite{D1}, \cite{D2},\cite{D5},\cite{X5},\cite{X8}. Real social networks, however, have much more complicated structure, that is better captured by special network models. There are a lot of candidates of social entwork models, such as the Chung-Lu model\cite{X1}, the Sznajd model\cite{X7}, the Barab\'{a}ási-Albert (BA) model \cite{C1} ,and the Generalized Markov Graph (GMG) model\cite{C10}. The GMG models can reveal intrinsic statistical properties of a social network (e.g., connectivity patterns, measured via distributions of centralities) and thus help analyze the information flow model.

Such intrinsic properties can (at least in theory) be exploited when one studies prediction or control problems in social networks. The properties of the network could be effectively used to determine, for example, which people should be chosen to spread the information to others in order to  minimize the number of such people, maximize the number of people with the desired behaviour in the network, which is novel compared to other IFM on social networks\cite{X6}, \cite{X7}. In another example, in a medical study with a limited budget, simulations of  the social network could be used to select the cohort of patients in the hope of an expedited and less costly study.


In this paper, we propose an IFM with enhanced (that is, BA or GMG) models for describing the social network properties.  Our motivation for proposing the specific IFM is the desire to design  strategies that can control beliefs effectively, without excessive overheads in computation time or memory requirements. 



The contributions of this paper are the following. First, we propose an IFM adapted to a realistic social network without excessive overheads. Second,  we develop three methods to analyse beliefs in a social network; the methods can be tuned to trade off accuracy for overhead. Third, we develop strategies to control beliefs. Partial results of the paper also appeared in a brief conference paper~\cite{X2}. In addition to the contents in the conference version, experiments to verify basic assumptions of models, techniques to estimate the converged beliefs as well as proofs for all the theorems and detailed description of models are added in this paper. 


 
The paper is organized as follows. In Section~\ref{IFM}, we  define the concepts of belief, control strategy and social network, and introduce the IFM formally. We introduce the notion of ``control power'' as a metric that can be used to compare control strategies. In Section~\ref{NetworkModel}, we elaborate on the BA and GMG models and show how one can analytically calculate the respective control power metric. In order to verify the proposed model, we use real network data in Section~\ref{expall}, to test the fundamental properties of the model, and specific control strategies. We finally conclude with some remarks and future planned work in Section~\ref{cfr}.

\section{Information Flow Model}
\label{IFM}

\subsection{Basic concepts in IFM}


The basic elements that comprise an IFM are: the social network, the belief and the control strategy.

\subsubsection{Social Network}
 For a social network $G$ with $N$ nodes, we use indices $i \in \{1,2,...,N\}$ to represent the nodes; the set of nodes is defined as: $node_G = \{1,2,...,N\}$. The set of edges, $edge_G$, includes all pairs of connected nodes in the network, $\{i,j\},$ where $i \in node_G, j \in node_G$;  the social network is thus defined as: $G= \{node_G,edge_G\}$. 
 The social network can also be represented by its adjacency matrix $\textbf{A}$, whose elements are $A_{ij}$, where $A_{ij}=1$ if $\{i,j\} \in edge_G$; $A_{ij} = 0$ otherwise.

\subsubsection{Belief}

We employ two kinds of beliefs in the IFM model: private   and updated belief. The former is unchanged and taken as an input of the model. The latter, however, is updated at each time step. 

\paragraph{Private belief}
Private beliefs abstract the intrinsic characteristics of nodes in a network. They will not be changed during the process of information flow. In this model, we assume all nodes in the network have same probabilistic distribution of private beliefs, which means node $i$ in the network takes the private belief as a random number $w_i \in [-1,1]$ with distribution $p(w_i)$. And we denote the private belief vector as $\textbf{w}$, whose elements are $w_i$. The distribution $p(\textbf{w})$ is common knowledge to everyone in the network.

\paragraph{Current belief}
A current belief $B_{i,T}$ describes the current opinion of node $i$ in a network at time step $T\in\mathbb{N}$. It lies in the range $[-1, 1]$.   $\mathbf{B}(T)$ is a vector whose elements are $B_{i,T}$. $\mathbf{B}(T)$ will be updated at each time step, and will converge to a limit $\mathbf{B}(T)$ in certain networks, as will be explained in Section \ref{NetworkModel}.


\subsubsection{Control Strategy}
To control the overall behaviour of the network, we propose a control strategy which chooses certain nodes in the network, the so called control nodes, and asks them to broadcast certain beliefs to their neighbours. 

\paragraph{Control Set}
The set of $c$ control nodes is defined as $\mathbf{C} = \{\theta_1,\theta_2,...,\theta_c\}$, where $\theta_i$, $1\leq i \leq c$, are indices of control nodes. The uncontrolled nodes  thus belong to set $^{\dagger}C = \{\theta_{c+1},...,\theta_N\}$.  
And the belief chosen to be broadcast by the $i^{th}$ control node, is set to a controlled belief $B^{*}_i$, where $B^{*}_i \in [-1,1]$, such that:
\begin{equation}
w_{\theta_i} = B^{*}_i, B_{\theta_i,T} = B^{*}_i,
\end{equation} 
for $\theta_i \in \mathbf{C}$ and any value of $T$. A control strategy is specified by the control set $\mathbf{C}$ with the corresponding controlled beliefs $\mathbf{B^{*}}$.

{

\paragraph{Control Power}
 Control power is used to measure how much the beliefs in a network have changed from their initial status. Control power for an arbitrary node $i$ is defined as the expectation of the difference between the ``final'' belief $B_{i,\infty}$ and the initial belief $w_i$: 

\begin{equation}
cp_i = \mathbf{\mathit{E}}[{B_{i,\infty} - w_i}].
\end{equation}
The averaged $cp_i$ values over all nodes in the network is called the {\textit{network control power}}: 
\begin{equation}
\label{cp}
cp = \Sigma_{i=1}^{N} \mathbf{\mathit{E}}[B_{i,\infty} - w_i]/N.
\end{equation}


\subsection{Information Flow Model}
\label{IMFSUB}

In an information flow model, $B_{i,0}$ is initialized to $w_i$. The value of $B_{i,T}$ is then updated at each time step and determined as the average of the current beliefs of the neighbours of node $i$, and the private belief of node $i$, $w_i$, under the influence of a control strategy. The average process is specified by an adjusted private belief vector and an adjusted adjacency matrix. The adjusted private belief vector is denoted by $\mathbf{w}^{*}$, with elements $w_i/(d_i+1)$, where $d_i = \Sigma_{j=1}^{N}A_{ij}$. And the adjusted adjacency matrix $\textbf{A}^{*}$ contains elements $A^{*}_{i,j} = A_{i,j}/(1+d_j)$. The control strategy is specified by a control matrix and a control vector, which are determined by the control set $\mathbf{C}$ and the corresponding controlled belief $\mathbf{B^*}$. The control matrix is defined as $\textbf{M}$, where $M_{i,i} = 1$ if $i\notin \mathbf{C}$ and $M_{i,j} = 0$ otherwise. The control vector is $\textbf{V}$, where $V_{\theta_i} = B^{*}_i$ if $i \leq c$, and $V_{\theta_i} = 0$ otherwise. If $\mathbf{C} = \emptyset$, the control strategy is trivial. The updating process of the current belief vector $\mathbf{B}(T)$ is shown in Figure (\ref{USYS}).

\begin{figure}[htb]
\centering
\centerline{\includegraphics[width=9.5cm]{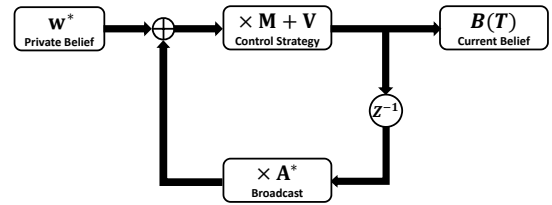}}
\vspace{-3.5cm}
\caption{Process for Information Update.}
\label{USYS}
\end{figure}
 
 In Figure (\ref{USYS}), the adjusted private belief $\mathbf{w}^{*}$ and adjusted adjacency matrix $\textbf{A}^{*}$ are for calculating an averaged belief for each node. The control matrix $\textbf{M}$ and control vector $\textbf{V}$ are used to set the belief of control nodes as their corresponding controlled belief $\mathbf{B^*}$. And $Z^{-1}$ means the beliefs are updated based on the information of the previous time step. Equation (\ref{CBCCS}) shows the formula to calculate $\mathbf{B}(T)$: 

\begin{equation}
\label{CBC}
\mathbf{B}(T) = [\mathbf{w}^{*} + \mathbf{B}(T-1) \times A^{*}]\times M + V.
\end{equation}


 Based on Equation (\ref{CBC}), we can derive a compact sigma notation of current belief as shown in Equation (\ref{CBCCS}):

\begin{equation}
\label{CBCCS}
\mathbf{B}(T) = [\mathbf{w}^{*} \times M + V]\times[ \Sigma_{t_1 =0}^{T-1} (A^{*}\times M)^{t_1}].
\end{equation}

assuming that the summation $\Sigma_{t_1=0}^{T} (A^{*}\times M)^{t_1}$ in Equation (\ref{CBCCS}) converges to a finite matrix when $T$ approaches $\infty$, the converged belief $\mathbf{B}(\infty)$ can be represented as:

\begin{equation}
\label{CBCCF}
\mathbf{B}(\infty) = [\mathbf{w}^{*} \times M + V]\times[ I - A^{*}\times M]^{-1}.
\end{equation}


{From Equation (\ref{CBCCF}) and Equation (\ref{cp}), we obtain the   control power metric when using a control strategy with parameters $\textbf{M}$ and   $\textbf{V}$ as\footnote[1]{$||\mathbf{x}||_1$ is the $\mathit{l_1 norm}$ of vector $\mathbf{x}$ with N elements: $||\mathbf{x}||_1 = \Sigma_{i = 1}^{N}x_i$}: 

\begin{equation}
\label{cpeq}
cp = \frac{1}{N}||\mathbf{\mathit{E}} \bigg[ [\mathbf{w}^{*} \times M + V]\times[ I - A^{*}\times M]^{-1} - \textbf{w} \bigg]||_1.
\end{equation}
}


\subsection{An example}
 In certain applications, a specific direction of change of public beliefs may be of interest, and hence a desire to control the sign of $cp$, and maximize its magnitude. For example, in a social network \textit{G}    the set $node_G$ may represent   patients with certain disease, and the set $edge_G$ may be  determined by the email and phone communications between these patients. Each patient has his/her own private belief $w_i$ about their disease. A doctor may want to influence the beliefs ${B(T)}_i$ of his/her patients and steer their final beliefs ${B(\infty)}_i$  towards a desired belief (e.g.,  ``following your  suggested treatment will help you''). To do so, the doctor may choose several patients in the social network as control nodes $\mathbf{C}$, and ask them to broadcast certain controlled beliefs $\mathbf{B^*}$ about taking the suggested treatment. However, the number of controlled patients $c$ may be limited by budget constraints. To reach  maximum performance regarding influencing the patients beliefs, the doctor needs an optimized control strategy, as we describe later. 

\section{Analysis of Control Power}
\label{NetworkModel}

In this section, we show how we can efficiently calculate the measure of control power in Equation~(\ref{cpeq}) and select an appropriate control strategy for social networks via two social network models, the BA model and the GMG model, each of which has its own advantage in performance or resources in computation and information. We start each model by introducing the fundamental assumptions and the network synthesis processes, which will be used to verify the assumptions. The main results regarding calculation of control power are Equation (\ref{cpPAeq}) and Equation (\ref{cpGeq}). Theorem~\ref{ConT1} and ~\ref{ConT2} are the main results regarding optimization of the control power, as they provide optimized control strategies for each model.

According to Equation (\ref{cpeq}), to calculate the control power $cp$, the complete information about $\textbf{A}$ is needed. In addition, the computational cost is the inversion of matrix $I - \textbf{A}^{*}\times \textbf{M}$. Such an exact solution does not shed any particular  light on the choice of control set $\mathbf{C}$, or on the convergence speed of $\mathbf{B}(T)$ towards $\mathbf{B}(\infty)$. In order to reduce the information needed to predict the control power $cp$, as well as provide a detailed analysis about the control strategy and convergence speed, network models are required. In particular, the social network models are meant to reveal the intrinsic properties of $\textbf{A}$, as the other elements in the proposed information flow model are well understood.
 

In practice,  information about the network may not be complete, which means $\textbf{A}$ is not always available. Furthermore, the network may contain a large number of nodes or edges, which requires significant computational power to process. To solve such problems, network models are necessary. A good network model can help calculate the converged beliefs using less information than $\textbf{A}$, and more efficiently. Two important network models, the BA model\cite{C1} and the GMG model\cite{C10}, will be introduced and applied in the analysis of the information flow model. The reason for choosing these two models is that they both provide probabilistic properties about the element $A_{i,j}$ of the adjacency matrix $\textbf{A}$. The BA model assumes that $A_{i,j}$ only depends on the degree of nodes $i$ and $j$, and thus requires less information. The GMG model, on the other hand, extends the dependence of $A_{i,j}$ to both degree and clustering coefficient of nodes $i$ and $j$, and thus provides better accuracy.

}
\subsection{Barab\'{a}ási-Albert Model}
\subsubsection{Basic assumption}
\label{PAMASS}
{A BA model describes the growth of a network. We can, however, model a static network of interest with size $N$ as one evolves from an initial network of small size until the the number of nodes reaches $N$. Then the network growth is freeze and we perform information flow on it. The purpose of introducing the BA model is to reveal the statistical characteristic of the adjacency matrix of a social network. The BA model, however, receives different judgements from academics\cite{X3}\cite{X4}, which leaves room for improvement such as the proposed GMG model.

One of the basic assumptions of the BA model is that the probability of a node $i$ attached by a new edge is proportional to its degree $d_i$\cite{C1}:

\begin{equation}
\label{PA}
\frac{\partial d_i}{\partial t} \sim d_i.
\end{equation}

Based on this assumption, we can derive the probability $P_{i,j}$ of an edge established between nodes $i$ and $j$ when the size of the network grows to a certain size, as shown in Theorem ~\ref{AijPT}. In this information flow model, $A_{i,j}$ is binary, so that $^{1}P_{i,j}$ is also the expected value of $A_{i,j}$ as: $\overline{A}_{i,j} = 1\times ^{1}P_{i,j} + 0\times(1-^{1}P_{i,j})$. Moreover, the summation of $\overline{A}_{i,j} = P_{i,j}$ over all choices for $i$ and $j$ is $\Sigma_{k=1}^{N}  d_k$, which is the summation of all degrees and is $\Sigma_{i,j}  A_{i,j}$. $^{1}P_{i,j}$  plays an important role in the analysis of control power estimation and of control strategy, as it represents the information about the matrix $\textbf{A}$.

\newtheorem{thm}{Theorem}[section]
\begin{thm}
\label{AijPT}
In a BA model, the probability $^{1}P_{i,j}$ of two nodes $i$ and $j$ being connected in network $G$ with $N$ nodes and a fixed degree sequence is:
\begin{equation}
\label{AijP}
^{1}{P}_{i,j} = \frac{d_i d_j}{\Sigma_{k=1}^{N}  d_k},
\end{equation}
where $d_k$ is the degree corresponding to node $k$. 
\end{thm}
The proof of Theorem ~\ref{AijPT} is provided in Appendix~\ref{App1}.


\subsubsection{Network Synthesis}
\label{synpa}
 
In order to verify the correctness of Equation (\ref{AijP}), we need to generate sample social networks according to the basic assumption of the BA model. The network synthesis of a BA model is discussed in Barab\'{a}ási(2002)\cite{C1}. The input of the synthesis process is the total number of nodes $N$ and the average number of edges attached to each new incoming node $m$. The detailed process is shown in Appendix~\ref{pro1}.

\subsubsection{Calculation of Control Power}
\label{cppa}
{
Theorem~\ref{AijPT} has revealed the statistical properties of adjacency matrix $\textbf{A}$. If we take $\textbf{A}$ as a random matrix, combined with the formula to calculate a converged belief, as shown in Equation (\ref{CBCCS}) and Equation (\ref{CBCCF}), we are able to obtain the expected value of converged beliefs for all the nodes in the network, which is shown in Theorem~\ref{Thm2}.

\begin{thm}
\label{Thm2}
In a BA model, the expected value of converged belief $^{1}B_{i,\infty}$ of a non-controlled node $i$, $i\not\in \mathbf{C}$, is:
\begin{equation}
\label{CBB}
\overline{^{1}B}_{i,\infty} = \frac{1}{\Sigma_{k=1}^{N}  d_k}\frac{d_i}{1+d_i} \frac{{\Sigma_{j=1}^{c}{B^*}_{\theta_j} d_{\theta_j}} + {\Sigma_{j=c+1}^{N}\frac{\overline{{w}}_{\theta_j}}{1+d_{\theta_j}} d_{\theta_j}}}{1-{\beta_1}},
\end{equation}

where ${w}_{\theta_i}$ is the private belief of node $\theta_i$, $m$ is the average number of edges per node in a network $G$ with $N$ nodes, $d_i$ is the degree corresponding to node $i$, $B^*_{\theta_j}$ is the controlled belief of control node $j$, $c$ is the number of control nodes, $\theta_i \in \mathbf{C}$ for $i\leq c$, and ${\beta_1}$ is a constant which is smaller than $1$:

\begin{equation}
\label{betaB}
{\beta_1} = {\Sigma_{k=c+1}^{N}\frac{{d_{\theta_k}}^{2}}{1+d_{\theta_k}}}/{\Sigma_{k=1}^{N}  d_k}.
\end{equation}

\end{thm}
}
The proof of Theorem~\ref{Thm2} is in Appendix~\ref{App2}.

Plugging Equation (\ref{CBB}) into Equation (\ref{cp}), we obtain the control power:

\begin{equation}
\label{cpPAeq}
{^{1}cp} = { \Sigma_{i=1}^{N} ({ \frac{{d_i}({{\Sigma_{j=1}^{c}B^*_{\theta_j} d_{\theta_j}} + {\Sigma_{j=c+1}^{N}\frac{\overline{{w}}_{\theta_j}}{1+d_{\theta_j}} d_{\theta_j}}})}{N\Sigma_{k=1}^{N}  d_k({1+d_i})({1-{\beta_1}})}  - \overline{w_i}} )}
\end{equation}
 

The information needed for the calculation in Equatoin(\ref{cpPAeq}) is the degree list of network $G$, which is far less than the information of adjacency matrix $\textbf{A}$. In addition, degree lists follow the power-law distribution in most social networks\cite{C1}, which means the degree list could be sampled from the network. The computational cost of such calculation is $O(N)$, which is much more efficient than the matrix inverse calculation required by Equation (\ref{cpeq}). 

\subsubsection{Optimization of Control Strategy}

According to Equation (\ref{cpPAeq}), we can see that the control strategy, as well as the degrees of the control nodes, have a direct impact on the control power. Without loss of generality, we set the preferred sign of beliefs positive. If controlled beliefs $B^*_i$, $i \leq c$, are maximized to be 1, and the private belief $w_i$ has zero mean, then, as shown in Theorem~\ref{ConT1},   the maximization of ${^{1}cp}$ requires the selection of a control group $\mathbf{C}$ to include nodes with highest degrees in the network $G$.

\begin{thm}
\label{ConT1}
Consider a social network G of $N$ nodes, with degree list $\{d_i\}$, $i = 1,\dots, N$, a private belief $\mathbf{w}$ with zero mean and  maximized control beliefs $B^*_i = 1$, $i \leq c$. Suppose further that the number of control nodes, $c$, is fixed.  The control set $\mathbf{C_o} = \{ {\theta_o}_1,{\theta_o}_2, \dots, {\theta_o}_c \}$, where $d_{{\theta_o}_i} \geq d_{{\theta_o}_j}$ if $i\leq j$, $1\leq i,j \leq N$, maximizes the   control power $^1 cp$:

\begin{equation}
\mathbf{C_o} = \arg\max_{\mathbf{C}} {^1 cp(\mathbf{C})}
\end{equation}

\end{thm}
The proof of Theorem~\ref{ConT1} is provided in Appendix~\ref{App4}.

\subsection{Generalized Markov Graph Model}

\subsubsection{Basic assumption}
\label{GMGASS}
{In  \cite{C10}, the BA model is shown to be a special case of a Markov Graph model\cite{C8}. A Markov Graph model is based on the dependence between pairs of nodes. The basic assumption of the BA model depends however on the degree of nodes, which is a specific description of pairwise relationship between nodes. It is thus natural to extend the BA model to the GMG model to analyse the property of adjacency matrix $\textbf{A}$. }

{In a GMG Model, the probabilistic dependence of an edge is extended from the other attached edges to attached triangles. As degree is used to describe the dependence on attached edges, a clustering coefficient, which is related to both edges and triad relational structures, is added to the description of dependence in a GMG model. The assumption about the probability of a node $i$ attached by a new edge in a GMG model then becomes: 

\begin{equation}
\label{PAG}
\frac{\partial d_i}{\partial t} \sim d_i (1+{\gamma}_i)^{\alpha},
\end{equation}
where $d_i$ is the degree of node $i$, $\gamma_i$ is the clustering coefficient of node $i$, and $\alpha$, which is called the clustering weight, is determined by the property of the network $G$.
}

{
The range of the clustering coefficient lies in $[0,1]$. If there is no triangle attached to a node $i$, the clustering coefficient $\gamma_i$ will be zero. To ensure a non-zero probability of a node getting an edge, on account of a zero clustering coefficient, we adopt a $(1+{\gamma}_i)$ in our model.
}

{
In an arbitrary network $G$, the influence of degree $d_i$ and clustering coefficient $\gamma_i$ on the probability of node $i$ obtaining a new edge is not equivalent. Parameter $\alpha$ is thus used to adjust the importance of clustering coefficient versus degree. The value of $\alpha$ changes for different types of networks and different applications.
}

{
Based on Theorem~\ref{AijPT}, we add the influence of clustering coefficients, and to make the summation of $^{2}P_{i,j}$ still be the sum of degrees $\Sigma_{k=1}^{N}  d_k$, the probability $^{2}P_{i,j}$ becomes:

\begin{equation}
\label{APij}
\begin{split}
&^{2}{P}_{i,j}= \frac{d_i (1+\gamma_i)^{\alpha} d_j (1+\gamma_j)^{\alpha}}{\eta}\Sigma_{k=1}^{N}  d_k,\\
\end{split}
\end{equation}

where $\eta$ is:
$$\eta = \Sigma_{i=1}^{N}\Sigma_{j=1,j\neq i}^{N} d_i (1+\gamma_i)^{\alpha} d_j (1+\gamma_j)^{\alpha}.$$
}

\subsubsection{Network Synthesis}
\label{syngm}

In order to verify the correctness of Equation (\ref{APij}), we need to generate sample social networks according to the basic assumption of the GMG model. In addition, the synthesized social networks can be used to estimate the clustering weight $\alpha$. In a GMG model\cite{C12}, the input of the synthesis process is: the total number of nodes $N$, the average number of edges attached to each node $m$, and the clustering weight $\alpha$. The detailed process is shown in Appendix~\ref{pro2}.

\subsubsection{Calculation of Control Power}
{Since the GMG model makes use of more information than the BA model, the probability $^{2}P_{i,j}$ should be more accurately describing the statistical property of adjacency matrix $\textbf{A}$ than $^{1}P_{i,j}$. Based on this idea, we develop the converged belief for a GMG model, as shown in Theorem~\ref{ThmG}. 

\begin{thm}
\label{ThmG}
Define a constant $\beta_2$ as,
\begin{equation}
\label{betaB2}
{\beta_2} = {\Sigma_{k=c+1}^{N}\frac{({{d_{\theta_k}(1+\gamma_{\theta_k})^{\alpha}}})^{2}}{(1+d_{\theta_k}){\eta}}}{\Sigma_{k=1}^{N}  d_k}.
\end{equation}

If $|\beta_2| < 1$, the expected value of converged belief $^{2}B_{i,\infty}$ of a non-controlled nodes $i$, $i\not\in \mathbf{C}$, in a GMG model, is,
\begin{equation}
\label{Geq}
\begin{split}
\overline{^{2}B}_{i,\infty} = &\frac{\Sigma_{k=1}^{N}  d_k}{\eta}\frac{d_i(1+\gamma_i)^{\alpha}}{1+d_i} \\
&\frac{{\Sigma_{j=1}^{c}B^*_{\theta_j} {d_{\theta_j}(1+\gamma_{\theta_j})^{\alpha}}} + {\Sigma_{j=c+1}^{N}\frac{\overline{{w}}_{\theta_j}}{1+d_{\theta_j}} {d_{\theta_j}(1+\gamma_{\theta_j})^{\alpha}}}}{1-{\beta_2}},
\end{split}
\end{equation}

where ${w}_j$ is the private belief of node $i$ in a network $G$ with $N$ nodes, $d_i$ is the degree corresponding to node $i$, $\gamma_i$ is the clustering coefficient of node $i$, $B^*_{\theta_j}$ is the controlled belief of control node $j$, $c$ is the number of control nodes, $\theta_j \in \mathbf{C}$ if $j \leq c$, $\alpha$ is the clustering weight for network $G$, $\eta$ is defined in Equation (\ref{APij}).
\end{thm}

The proof of Theorem~\ref{ThmG}} is in Appendix~\ref{App3}.

The constant $\beta_2$, however, will not be guaranteed to be strictly smaller than $1$, as $\beta_1$ is in a BA model, which is shown in Appendix~\ref{App2}. The value of $\beta_2$ will be determined by the degree list, the clustering coefficient list and the clustering weight $\alpha$ of the network $G$.

Plugging Equation (\ref{Geq}) into Equation (\ref{cp}), we obtain the control power:

\begin{equation}
\label{cpGeq}
\begin{split}
&{^{2}cp} = \frac{1}{N}\Sigma_{i=1}^{N}(\frac{\Sigma_{k=1}^{N}  d_k}{\eta}\frac{d_i(1+\gamma_i)^{\alpha}}{1+d_i}\times \\
&\frac{{\Sigma_{j=1}^{c}B^*_{\theta_j} {d_{\theta_j}(1+\gamma_{\theta_j})^{\alpha}}} + {\Sigma_{j=c+1}^{N}\frac{\overline{{w}}_{\theta_j}}{1+d_{\theta_j}} {d_{\theta_j}(1+\gamma_{\theta_j})^{\alpha}}}}{1-{\beta_2}} - \overline{w_i}).
\end{split}
\end{equation}


The calculation of a control power in Equation(\ref{cpGeq}) requires the information of the degree list, the clustering coefficient list and the clustering weight $\alpha$ of network $G$. The clustering weight $\alpha$ is obtained by a learning process, and is introduced in Section~\ref{expall}. When calculating the control power, the information needed by the GMG model is still far less than the information of adjacency matrix $\textbf{A}$. And due to the fact that $\eta$ could be rewritten as:
$$ {\eta} = (\Sigma_{i=1}^{N} {d_i(1+\gamma_i)^{\alpha}})^2 - \Sigma_{i=1}^{N} {d_i(1+\gamma_i)^{\alpha}},$$
the computational cost of such a calculation is $O(N)$, which is the same as that in the BA model. 

\subsubsection{Optimization of Control Strategy}

According to Equation (\ref{cpGeq}), we can see that, in addition to the control strategy and the degrees of the control nodes, the clustering coefficients of the control nodes also have a direct impact on the control power. Without loss of generality, we set the preferred sign of beliefs positive. If controlled beliefs $B^*_i$, $i \leq c$, are maximized to be 1, and the private belief $w_i$ has a zero mean, then, as we show in Theorem~\ref{ConT2}, the maximization of ${^{2}cp}$ requires the selection of a control group $\mathbf{C}$ to include nodes with highest $d(1+\gamma)^{\alpha}$ values in the network $G$.
 
\begin{thm}
\label{ConT2}
Consider a social network G of $N$ nodes, with degree list $\{d_i\}$, clustering weight $\alpha$,   clustering coefficient list $\{\gamma_i\}$, $i = 1,\dots, N$, and maximized control beliefs $B^*_i = 1$, $i \leq c$. Suppose further that the private belief $\mathbf{w}$ has   zero mean and the  number of control nodes $c$ is fixed. Then,  the control set $\mathbf{^2C_o} = \{ {{\theta^{\dagger}_o}}_1,{{\theta^{\dagger}_o}}_2, \dots, {{\theta^{\dagger}_o}}_c \}$, where $d_{{\theta^{\dagger}_o}_i}(1+\gamma_{{\theta^{\dagger}_o}_i})^{\alpha} \geq d_{{\theta^{\dagger}_o}_j}(1+\gamma_{{\theta^{\dagger}_o}_j})^{\alpha}$ if $i\leq j$, $1\leq i,j \leq N$, maximizes the control power $^2 cp$

\begin{equation}
\mathbf{^2 C_o} = \arg\max_{\mathbf{C}} {^2 cp(\mathbf{C})},
\end{equation}

under the condition that:

\begin{equation}
\frac{1}{\beta_2} > 1 + max(1,2^{\alpha})\frac{\Sigma_{j=1}^{c} d_{{\theta^{\dagger}_o}_j}(1+\gamma_{{\theta^{\dagger}_o}_j})^{\alpha}}{\Sigma_{k = c+1}^{N} \frac{(d_{{\theta^{\dagger}_o}_k}(1+\gamma_{{\theta^{\dagger}_o}_k})^{\alpha})^2}{1+d_{{\theta^{\dagger}_o}_k}}}.
\end{equation}

\end{thm}
The proof of Theorem~\ref{ConT2} is provided in Appendix~\ref{App5}.

 The GMG model covers more types of social networks due to the presence of the clustering weight $\alpha$, an additional free parameter. Note that the range of $\alpha$ is   constrained only by the learning data set, not by the model. The specific condition in Theorem~\ref{ConT2} means that the optimized control strategy is appropriate only for certain types of social networks where the beliefs converge fast, as denoted   by $\beta_2$.  
 

\section{Experiments on Verification and Applications of information flow models}
\label{expall}

Three sets of experiments are described in this section. In the first set, presented in Section~\ref{ExpOne}, we verify the basic assumptions about $P_{i,j}$, the probabilistic nature  of the two network models, as showed in { Theorem~\ref{AijPT} and Equation~(\ref{APij}).} In the second set, presented in Section~\ref{cbe}, we test the performance  of the two social network models  when the private beliefs of nodes in a social network have   zero-mean.  
In Section~\ref{cso}, the last set of experiments compares the effects of control strategies on maximizing the control power of two different models.

Real network data\cite{C11} is used in all three types of experiments. There are $3$ different types of social networks: on-line social networks, p2p transmission networks and physicist collaboration networks. Each of these three different social networks includes several subtypes of networks. On-line social networks include Slashdot network data of August 2008 and of February 2009, Wiki-vote network data and Epinions network data. P2p transmission networks include Gnutella network data at five different times. Physicist collaboration networks include collaboration networks of physicists studying astrophysics, condensed Matter Physics, theoretical high-energy Physics, experimental high-energy Physics and general relativity. All of these effectively constitute $14$ subtypes of social networks and will be denoted by indices: $1,2,\dots,14$.  From each of these $14$ networks, we sampled 50 sub-networks using the same sampling method.

\subsection{Basic Assumption Verification}
\label{ExpOne}
The first type of experiments is designed to test the fundamental assumptions of the BA model, as shown in Section~\ref{PAMASS}, and GMG model, as shown in Section~\ref{GMGASS}. Both assumptions are tested on the synthesized data to show the consistence of the assumptions with the derived formula of $P_{i,j}$ in both models. The real network data are then used to test how $P_{i,j}$ computed in these two models fit into real applications. The performance of the models on synthesized data is better than that on real data, since the former were used to generate the data in the first place. The additional performance difference is due to data sampling noise as well as to the intrinsic unaccounted characteristics of the network itself. 
 
\subsubsection{Barab\'{a}ási-Albert model}
\paragraph{Verification of Theorem~\ref{AijPT} on synthetic data}
{In order to test the correctness of the basic assumption of the BA model, $^{1}P_{i,j}$ is computed to compare with $A_{i,j}$ averaged across 1,000 realizations of networks synthesized by the algorithm introduced in ~\ref{synpa}. An example of experimental results is shown in Figure (\ref{PAF}) with networks containing $100$ nodes and $300$ edges. The average number of edges per node is thus $m =3$. In this example, the average degree list is averaged across the degree lists of the 10,000 realizations of networks. The averaged degree list is then used to calculate $^{1}P_{i,j}$ as shown in Equation (\ref{AijP}). The experiment is then repeated for different values of $m$, and the relative error between $^{1}P_{i,j}$ and $\overline{A}_{i,j}$ averaged across all pairs of nodes is shown in Table~\ref{Tabpam}}.

\begin{figure}

\begin{center}
        \begin{subfigure}[b]{0.25\textwidth}
                \centering
                \includegraphics[width=\textwidth]{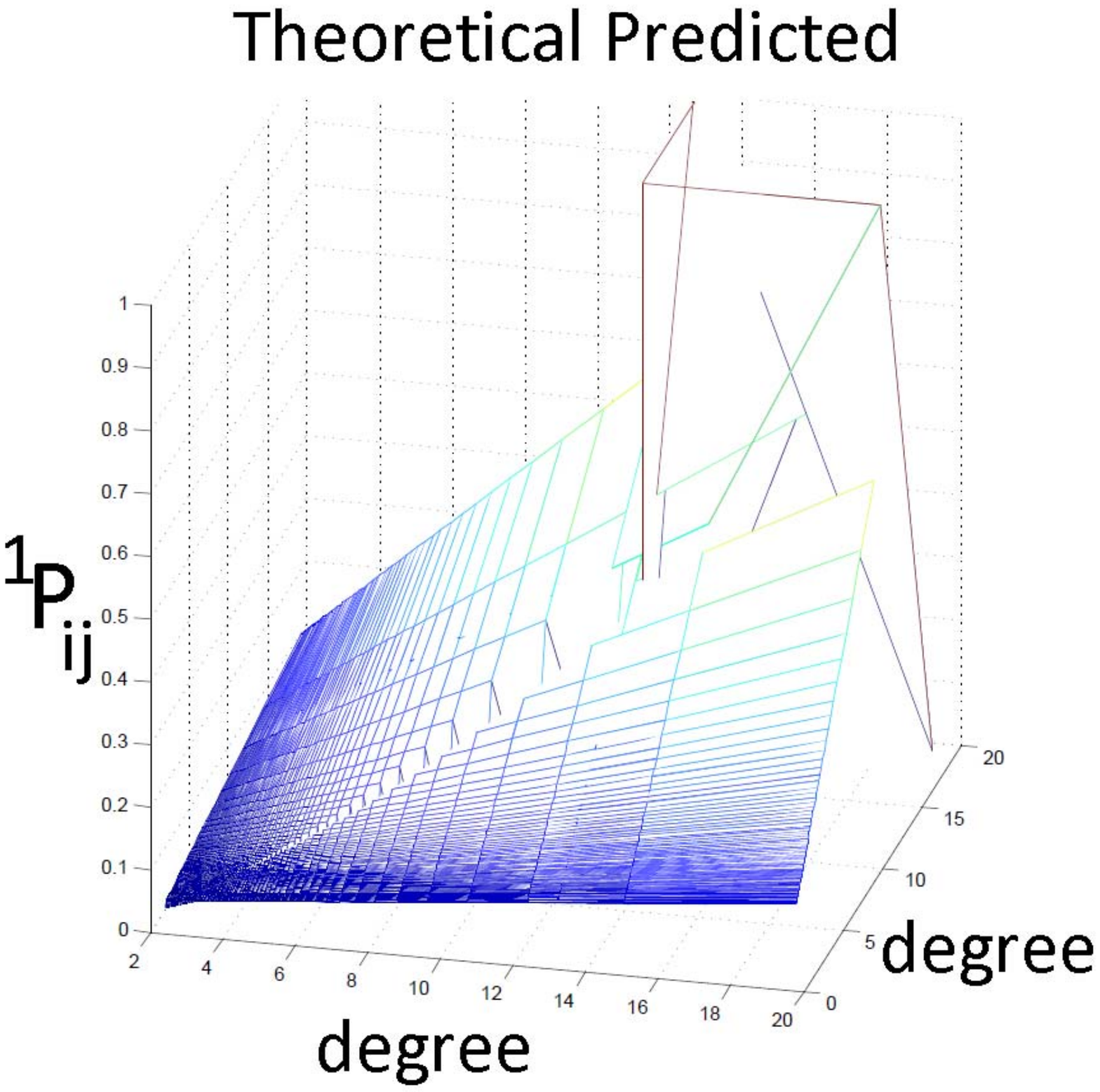}
                \caption{$^{1}P_{i,j}$}
        \end{subfigure}%
        \begin{subfigure}[b]{0.25\textwidth}
                \centering
                \includegraphics[width=\textwidth]{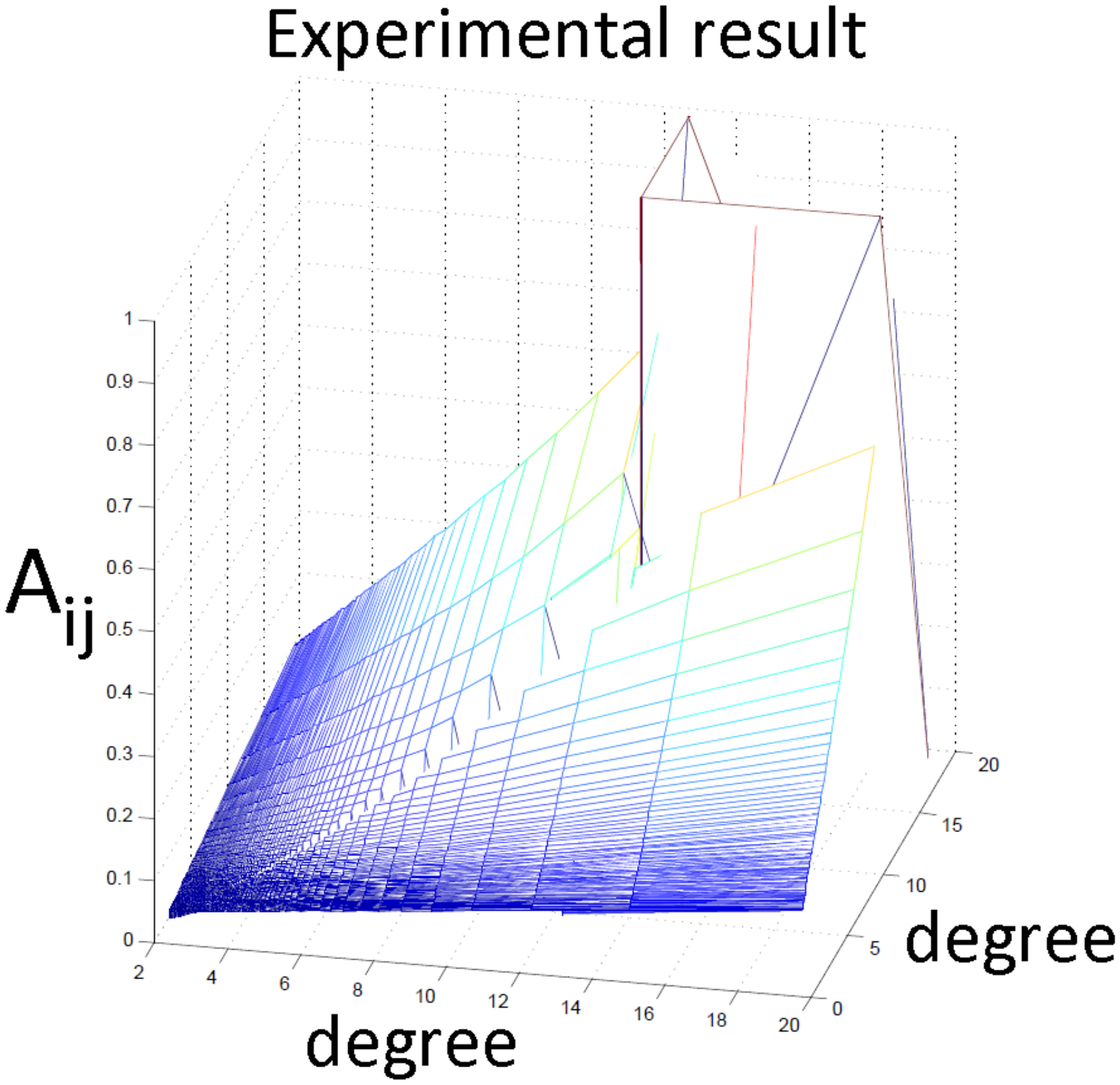}
                \caption{$\overline{A}_{i,j}$}
        \end{subfigure}
\end{center}

\caption{Comparison between $^{1}P_{i,j}$ and $\overline{A}_{i,j}$.}
\label{PAF}
\end{figure}

\begin{table}[!htbp]
\setlength{\tabcolsep}{5pt}
\centering
  \begin{tabular}{ |p{1cm} || c | c | c | c | c | }
    \hline
    m & 1 & 2 & 3 & 4 & 5 \\ \hline
    Relative Error & 4.32e-2 & 4.59e-2 &3.75e-2 &4.12e-2 & 3.61e-2 \\
    \hline
  \end{tabular}
  \caption{Relative error between $^{1}P_{i,j}$ and $\overline{A}_{i,j}$ on synthesized data.}
  \label{Tabpam}
\end{table}

From Table~\ref{Tabpam}, we can see that Equation (\ref{AijP}) can be used to calculate the value of $^{1}P_{i,j}$, which is the expected value of $A_{i,j}$ for different choices of $m$ value. This experiment thus verifies the result of Theorem~\ref{AijPT}.

\paragraph{Performance of Theorem~\ref{AijPT} on real network data}
{In this experiment, real network samples are used to validate Theorem~\ref{AijPT}. For each subcategory of network samples, the number of nodes $N$ is chosen to be the minimal number of nodes of all samples, which ensures the same number of samples being used for the calculation of the average number of every element in the adjacency matrix $A$. The degree list, $N$, and $m$ are then averaged across all samples of the same subcategory. The value of $N$, the averaged degree list and the averaged $m$ are plugged into Equation (\ref{AijP}) to calculate $^{1}P_{i,j}$. And the value of $\overline{A}_{i,j}$ is the average of $A_{i,j}$ from all 50 samples of networks belonging to the same subcategory. The relative errors between $^{1}P_{i,j}$ and $\overline{A}_{i,j}$ for 14 different types of network are shown in Table~\ref{Tabpar}.

\begin{table}[!htbp]
\centering
  \begin{tabular}{ |p{0.8cm} | p{1.1cm} || p{0.8cm} | p{1.3cm} ||  p{0.8cm} | p{1.1cm}  | }
    \hline
\multicolumn{2}{|c||}{On-line} & \multicolumn{2}{c||}{P2p} & \multicolumn{2}{c|}{Collaboration}\\
    \hline
    Net- work & Error & Net- work & Error & Net- work & Error\\ \hline
    1 & 8.31e-2 & 5 & 22.13e-2& 10& 7.28e-2\\ \hline
    2 & 8.57e-2 & 6 & 19.36e-2& 11& 9.15e-2\\ \hline
    3 & 7.46e-2 & 7 & 24.18e-2& 12& 7.16e-2\\ \hline
    4 & 6.59e-2 & 8 & 22.40e-2& 13& 8.52e-2\\ \hline
      &         & 9 & 20.25e-2& 14& 9.73e-2\\ \hline

  \end{tabular}
  \caption{Relative error between $^{1}P_{i,j}$ and $\overline{A}_{i,j}$ on real network data.}
  \label{Tabpar}
\end{table}

From Table~\ref{Tabpar}, we can see that the value of $^{1}P_{i,j}$, according to Equation (\ref{AijP}), is close to the expected value of $A_{i,j}$ in on-line social network samples and Collaboration networks. However, in p2p transmission networks, Theorem~\ref{AijPT} doesn't provide a good estimate of $\overline{A}_{i,j}$. The main reason for the inaccurate
estimation of $A_{i,j}$ in this case is that the degree distribution of such a network does not strictly follow the power law distribution, which means the structure of these networks is different from what the BA model can describe. This indicates that a better model is needed for a more accurate description of the real network data.

\subsubsection{Generalized Markov Graph Model}
\paragraph{Verification of basic assumption on synthetic data}
{The basic assumption of the GMG model is tested as follows. The number of nodes $N$ is $100$. Five different values of $m$: $1,2,3,4,5$, and 4 different values of $\alpha$: $-2,-1,1,2$, are chosen as the inputs of the network synthesis process introduced in ~\ref{syngm}. For each combination of $m$ and $\alpha$, $^{2}P_{i,j}$ is calculated and compared with $A_{i,j}$ averaged across 1,000 realizations of synthesized networks. The relative error between $^{2}P_{i,j}$ and $\overline{A}_{i,j}$ averaged across all pairs of nodes is shown in Table~\ref{Tabgmm}}.

\begin{table}[!htbp]
\centering
  \begin{tabular}{ |c || c | c | c | c | c | }
    \hline
    \backslashbox[0.2cm]{$\alpha$\kern-1.5em}{\kern-1em m} & 1 & 2 & 3 & 4 & 5 \\ \hline
    -2 & 3.25e-2 & 4.82e-2 &4.26e-2 &3.19e-2 & 3.55e-2 \\
    \hline
    -1 & 4.09e-2 & 4.91e-2 &4.92e-2 &3.31e-2 & 4.94e-2 \\
    \hline
    1 & 4.91e-2 & 3.97e-2 &4.60e-2 &3.28e-2 & 3.84e-2 \\
    \hline
    2 & 4.83e-2 & 4.58e-2 &4.91e-2 &4.31e-2 & 3.07e-2 \\
    \hline

  \end{tabular}
  \caption{Relative error between $^{2}P_{i,j}$ and $\overline{A}_{i,j}$ on synthetic data.}
  \label{Tabgmm}
\end{table}

From Table~\ref{Tabgmm}, we can see that Equation (\ref{APij}) can be used to calculate the value of $^{2}P_{i,j}$, which is the expected value of $A_{i,j}$ for different choices of $m$ and $\alpha$.

\paragraph{Verification of basic assumptions on real network data}
{In this experiment, real network samples are used to validate Equation (\ref{APij}). For each subcategory of the network samples, the number of nodes $N$ is chosen to be the minimal number of nodes of all $50$ samples. The degree list of length $N$, the clustering coefficient list of length $N$, and the average number of edges $m$ are then averaged across all samples of the same subcategory. Then 25 of randomly chosen sampled networks for each sub-category of networks are used as a training set to learn the value $\alpha$. The rest of sampled networks are used as a test set. For each sub-category of networks, $\alpha$ is determined as the value that generates $^{2}P_{i,j}$ closest to $A_{i,j}$ averaged across the same 25 chosen sampled networks. Next, for each sub-category of networks, averaged degree lists, average clustering coefficient lists and averaged $m$ of the other 25 sampled networks, along with the learned $\alpha$, are used in Equation (\ref{APij}) to compute $^{2}P_{i,j}$. The value of $\overline{A}_{i,j}$ is the average of $A_{i,j}$ over the test set for each subcategory of networks. The relative error between $^{2}P_{i,j}$ and $\overline{A}_{i,j}$ for 14 different types of network is shown in Table~\ref{Tabgmr}.

\begin{table}[!htbp]
  \begin{tabular}{ |p{0.8cm} | p{1.1cm} || p{0.8cm} | p{1.1cm} ||  p{0.8cm} | p{1.1cm}  | }
    \hline
\multicolumn{2}{|c||}{On-line} & \multicolumn{2}{c||}{P2p} & \multicolumn{2}{c|}{Collaboration}\\
    \hline
    Net- work & Error & Net- work & Error & Net- work & Error\\ \hline
    1 & 5.14e-2 & 5 & 8.68e-2& 10& 5.45e-2\\ \hline
    2 & 6.23e-2 & 6 & 7.51e-2& 11& 6.69e-2\\ \hline
    3 & 4.75e-2 & 7 & 8.64e-2& 12& 6.28e-2\\ \hline
    4 & 5.86e-2 & 8 & 9.13e-2& 13& 4.19e-2\\ \hline
      &         & 9 & 7.26e-2& 14& 6.94e-2\\ \hline

  \end{tabular}
  \caption{Relative error between $^{2}P_{i,j}$ and $\overline{A}_{i,j}$ on real network data.}
  \label{Tabgmr}
\end{table}

From Table~\ref{Tabgmr}, we can see that the value of $^{2}P_{i,j}$ from Equation (\ref{AijP}) is close to the expected value of $A_{i,j}$ for all three different categories of networks. Although in p2p transmission networks, Equation (\ref{AijP}) doesn't give as good an estimate of $\overline{A}_{i,j}$ as in the other two categories, the overall performance of the GMG model is better than that of the BA model. Note for p2p transmission networks, whose degree distribution does not strictly follow the power law distribution, the GMG model could greatly increase the accuracy of $A_{i,j}$ estimate.

\subsection{Converged Belief Estimation}
\label{cbe}
In this experiment, we evaluate the accuracy of the converged belief $\mathbf{B}(\infty)$ estimate for both models of interest. The control strategy is set to push neutral public opinions towards positive opinions. The private beliefs $w_i$ of nodes in the network are set to be neutral, i.e, they obey a uniform distribution on $[-1, 1]$. For both the BA and the GMG models, 25 randomly chosen networks are selected from each sub-category of networks, and used as a test set. The other 25 networks for each sub-category of networks are used as a training set for the GMG model to learn the clustering weight $\alpha$. 

\subsubsection{Barab\'{a}ási-Albert Model}
For each of the network samples in the test set of each sub-category of networks, a degree list is recorded. The control set $\mathbf{C}$ is set to be the top $5\%$ nodes with the highest degree in the network, thus fixing the number of control nodes to: $c = \lceil 5\%N\rceil$. The controlled beliefs $\mathbf{B^*}$ for nodes in the control set are set to be $1$. For each network sample, the private belief list ${w_i}$ is generated $100$ times according to a uniform distribution on $[-1, 1]$.  

For each of the generated private belief lists, the   value of control power ${^{1}cp_i}$ is calculated as in Section~\ref{cppa}. The same network is used for calculating the exact value of the control power according to Equation (\ref{CBCCF}). The relative error between the ${^{1}cp_i}$ and the exact value is then recorded as the relative error for this network sample under this belief list. The $100$ relative errors for all generated private belief lists are then averaged and recorded as the relative error of this network sample. Next, the relative errors for all network samples in the test set of each sub-category of networks are averaged and recorded as the relative error of the BA model on this sub-category. The relative errors for 14 different sub-categories of networks are shown in Figure (\ref{cppaG1}) and Figure (\ref{cppaG2}) to compare with the results from the GMG model.

\subsubsection{Generalized Markov Graph Model}
 
 In a GMG model, the clustering weight $\alpha$ plays a key role in describing the characteristics of the social network  and has to be learned from the training data. We propose two ways to learn $\alpha$. The first one makes use of complete  information of the adjacency matrix $A$ of the social network and   provides better accuracy. The second one learns $\alpha$ using less information, namely only the degree list and clustering coefficient list of the social network. The learned $\alpha$ is then used to calculate the     control power and the results are compared with the one from the BA model. 

\paragraph{learning $\alpha$ from $\textbf{A}$}
\label{GMGA}

The first step in a GMG model is to learn the clustering weight $\alpha$ from the training sets for each sub-category of networks. For each training network sample, the degree list and clustering coefficient list are recorded. The control set $\mathbf{C}$ is set to be the top $5\%$ nodes with highest degree in the network, and thus the number of control nodes is set to be: $c = \lceil 5\%N\rceil$. The controlled beliefs $B^*_i$ for nodes in the control set are set to be $1$, and $100$ private belief lists sampled from a uniform distribution on $[-1, 1]$ are obtained.

The clustering weight $\alpha$ is chosen to minimize the difference between ${cp}$ and ${^{2}cp}$. For each network sample in the training set of each sub-category of networks, the control power ${^{2}cp_i}$ is calculated for each of the $100$ private belief lists with an arbitrary choice of $\alpha$. The exact solution of control power is also calculated for the $100$ generated private belief lists. The relative error between the ${^{2}cp_i}$ and the exact value of the same private belief list is then calculated, averaged across all generated private belief lists, and recorded as the relative error for this network sample. For all $25$ network samples in the training set, the relative errors are calculated and their average value is recorded as the relative error of this sub-category of networks. The clustering weight $\alpha$ for each sub-category of networks is chosen as the one that minimizes the relative error, as shown in Table~\ref{alpha1T}.
 
\begin{table}[!htbp]
\centering
  \begin{tabular}{ |c | c || c | c ||  c  |  c  | }
    \hline
\multicolumn{2}{|c||}{On-line} & \multicolumn{2}{c||}{P2p} & \multicolumn{2}{c|}{Collaboration}\\
    \hline
    Network & $\alpha$ & Network & $\alpha$ & Network & $\alpha$\\ \hline
    1 & -1.00 & 5 & -0.20 & 10& 1.48\\ \hline
    2 & -1.00 & 6 & 1.29 & 11& -2.57\\ \hline
    3 & -0.50 & 7 & 1.52 & 12& -2.57\\ \hline
    4 & -0.30 & 8 & 1.45 & 13& -2.84\\ \hline
      &      & 9 & 1.12 & 14& 0.52\\ \hline

  \end{tabular}
  \caption{$\alpha$ learned from complete information of $\textbf{A}$ from training set.}
  \label{alpha1T}
\end{table}

The learned $\alpha$ is subsequently used to test the performance of the GMG model on the test set. For each of the network samples in the test set, the degree list and the clustering coefficient list are recorded. The control set is similar to that in the training set. For each network sample, the private belief list ${w_i}$ is generated $100$ times according to a uniform distribution on $[-1, 1]$. The relative errors are calculated on the test set by the same method as that used on the training set. The relative errors for 14 different sub-categories of networks are shown in Figure (\ref{cppaG1}), together with the result from the BA model.

\begin{figure}[htb]
\centering
\centerline{\includegraphics[width=8.5cm]{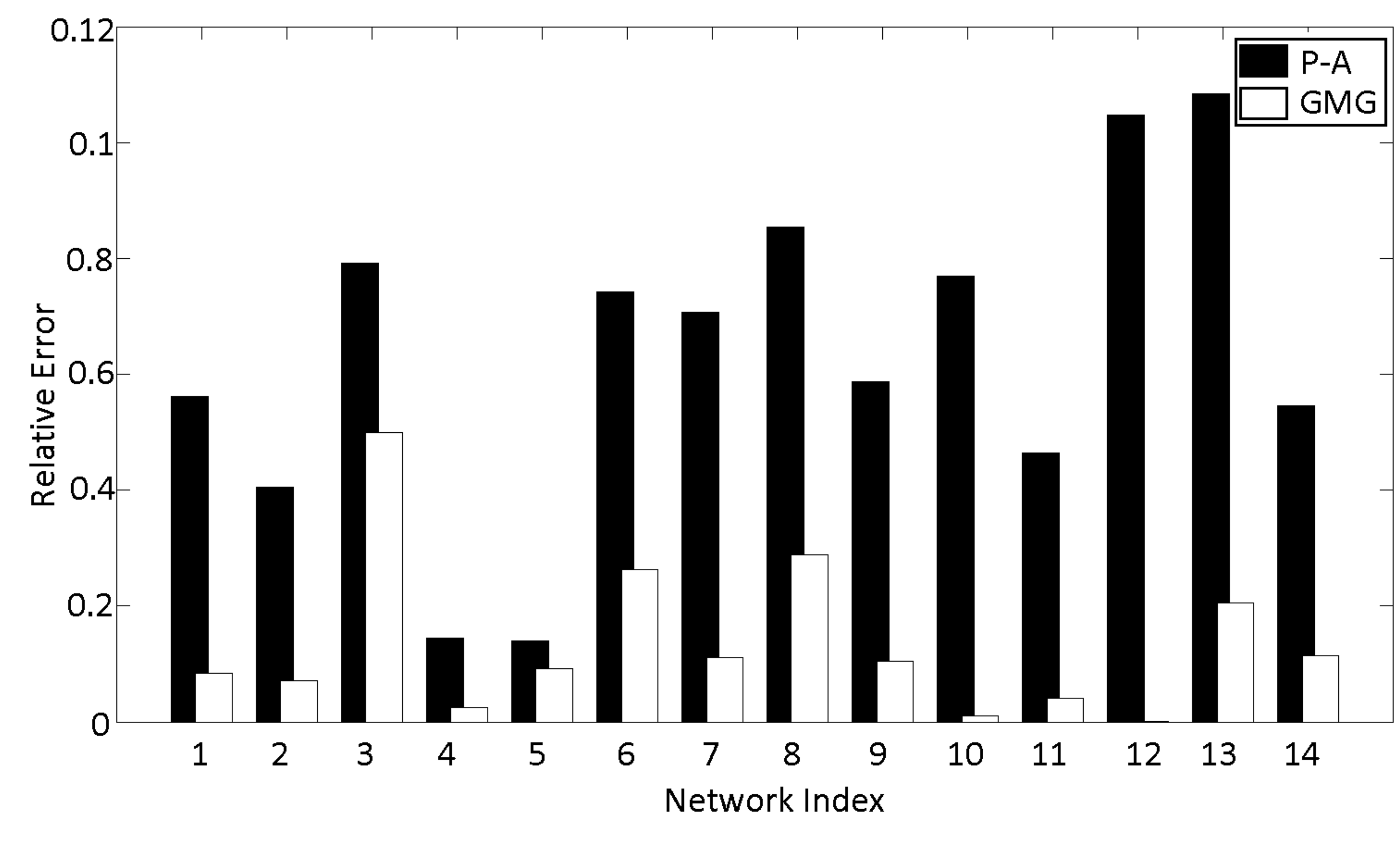}}
\caption{Relative error of control power}
\label{cppaG1}
\end{figure}

\paragraph{learning $\alpha$ from the degree list and the clustering coefficient list}

The second method of learning $\alpha$ only uses partial information of the networks in the training set, i.e, the degree list and the clustering coefficient list. For each of the network samples in the training set, the number of nodes $N$, the number of edges $N\times m$, the degree list and the clustering list are recorded. For an arbitrary value of $\alpha$, $100$ networks with $N$ nodes and $N\times m$ edges are generated according to the network synthesis algorithm introduced in Section~\ref{syngm}. For each of the generated networks, the degree list and the clustering coefficient list are recorded and averaged across these $100$ networks. The averaged degree list and clustering coefficient list are then compared with the one of the network sample in the training set. The difference between them is calculated for each of the network samples in the training set, and averaged across all $25$ networks in the training set. The $\alpha$ that minimizes the average difference is selected as the clustering weight of this sub-category of networks. The results for all $14$ sub-categories of networks are shown in Table~\ref{alpha2T}.

\begin{table}[!htbp]
\centering
  \begin{tabular}{ |c | c || c | c ||  c  |  c  | }
    \hline
\multicolumn{2}{|c||}{On-line} & \multicolumn{2}{c||}{P2p} & \multicolumn{2}{c|}{Collaboration}\\
    \hline
    Network & $\alpha$ & Network & $\alpha$ & Network & $\alpha$\\ \hline
    1 & -0.40 & 5 & -0.15 & 10& 0.58\\ \hline
    2 & -0.30 & 6 & 0.29 & 11& -2.30\\ \hline
    3 & -0.40 & 7 & 0.50 & 12& -1.50\\ \hline
    4 & -0.20 & 8 & 0.40 & 13& -1.70\\ \hline
      &       & 9 & 0.10 & 14& 0.50\\ \hline

  \end{tabular}
  \caption{$\alpha$ learned from partial information of $\textbf{A}$ from training set.}
  \label{alpha2T}
\end{table}

The learned $\alpha$ is then used to test the performance of the GMG model on the test set. The experiment is similar as in Section~\ref{GMGA}, and the relative errors of control power are computed for all $14$ subcategories of networks, as shown in Figure (\ref{cppaG2}).

\begin{figure}[htb]
\centering
\centerline{\includegraphics[width=8.5cm]{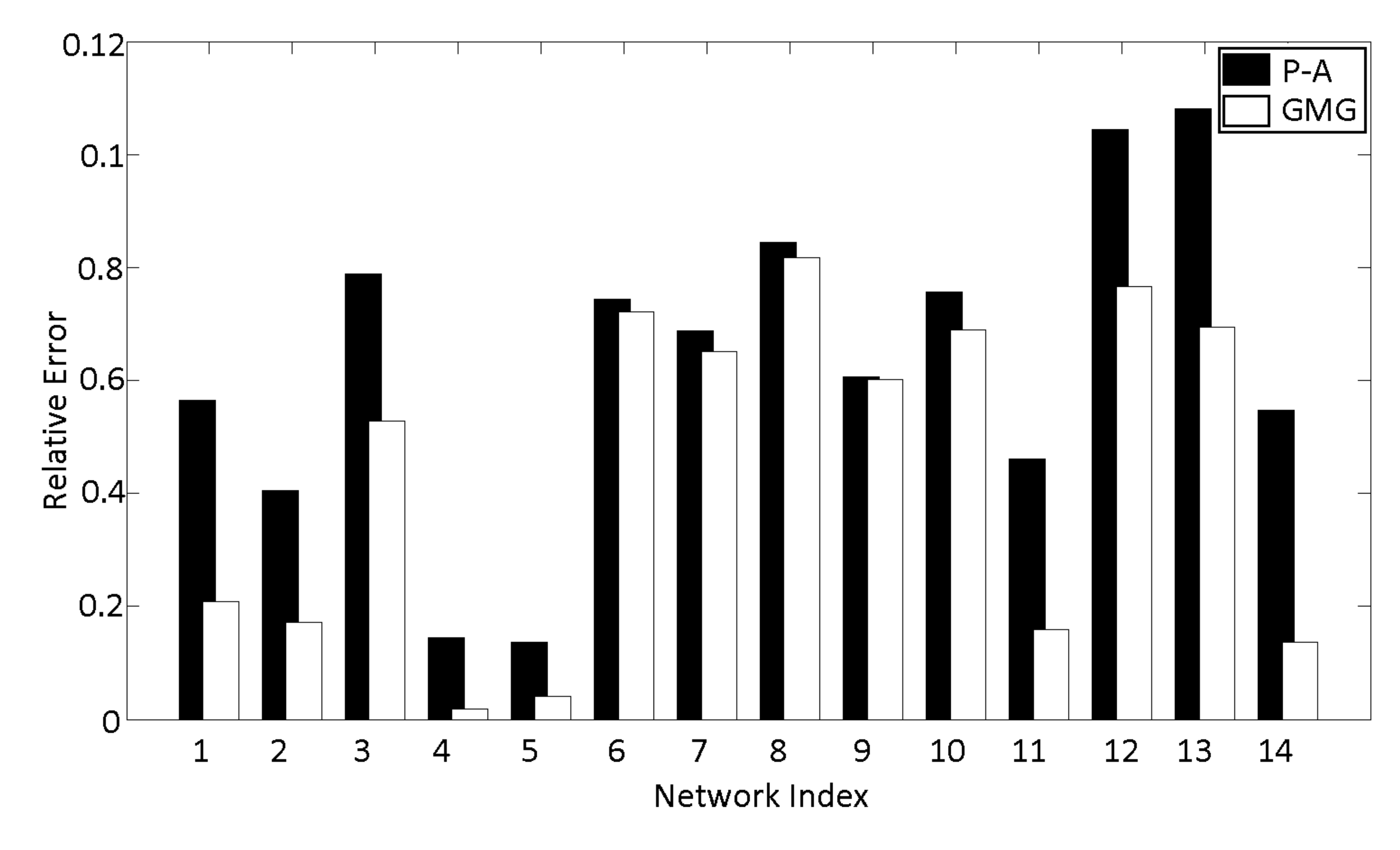}}
\caption{Relative error of control power}
\label{cppaG2}
\end{figure}

\paragraph{Discussion of the experiments}
 Figures (\ref{cppaG1}) and (\ref{cppaG2}) indicate that the GMG model outperformed the BA model, as the former one yields smaller error for both $\alpha$ learned from adjacency matrix $A$ and from the degree list and the clustering list. When the complete information of $\textbf{A}$ of the training set is used, the learned $\alpha$ will make the GMG model more accurate than the case when only partial information of $\textbf{A}$ is learned. And the more accurate of the models, the more information is needed.



 
\subsection{Control strategy optimization}
\label{cso}
In the experiments described in this section, we are interested in comparing the performance of control strategies under the two network models. 

\subsubsection{Barab\'{a}ási-Albert Model}
According to Theorem~\ref{ConT1}, the control strategy of the BA model requires the control set $\mathbf{C}$ to include the $c$ nodes with the highest degrees in the network. The test set is the same as in Section~\ref{cbe}. For each network in the test set, the adjacency matrices are recorded. The indices of the first $\lceil 5\%N\rceil$ nodes with the highest degrees are set as the first $\lceil 5\%N\rceil$ nodes, so that they constitute the control set. The private belief lists with uniform distribution on $[-1, 1]$ are sampled $100$ times.

For the networks in the test set of each sub-category network, the converged beliefs and control power of each network are calculated according to Equation (\ref{CBCCF}), based on each of the $100$ private belief lists. The control power averaged on these $100$ private belief lists and the $25$ networks is then recorded as the average control power of this sub-category of networks. The results are shown in Figure (\ref{str1}) and Figure (\ref{str2}).

\subsubsection{Generalized Markov Graph Model}

In a GMG model, according to Theorem~\ref{ConT2}, the control set should include nodes with highest $d(1+\gamma)^{\alpha}$ in order to achieve maximum control power. In this experiment, $\alpha$ is learned from either complete or partial information of the training set. The test set is the same as that in Section~\ref{cbe}. For each network in the test set, the adjacency matrix is recorded. The indices of the first $\lceil 5\%N\rceil$ nodes with highest $d(1+\gamma)^{\alpha}$ are selected for the control set. In order to utilize all the sample networks in the test set, the ones with $\beta_2$ values that do not satisfy the condition in Theorem~\ref{ConT2} are also included. Note that this will yield suboptimal control power values for a GMG model. However, as the experiments show, the control strategy from the GMG model still outperforms the best one from the BA model.

The private belief lists with uniform distribution on $[-1, 1]$ are sampled $100$ times. For each sub-category of networks, according to Equation (\ref{CBCCF}), the converged beliefs and control power of each network in the test set are calculated based on each of the $100$ private belief list for both $\alpha$ values as shown in Table~\ref{alpha1T} and Table~\ref{alpha2T}. The control power averaged across these $100$ private belief lists and the $25$ networks is recorded as the average control power of this sub-category of networks for each value of $\alpha$. The results are shown in Figure (\ref{str1}) and Figure (\ref{str2}).
 
\begin{figure}[htb]
\centering
\centerline{\includegraphics[width=8.5cm]{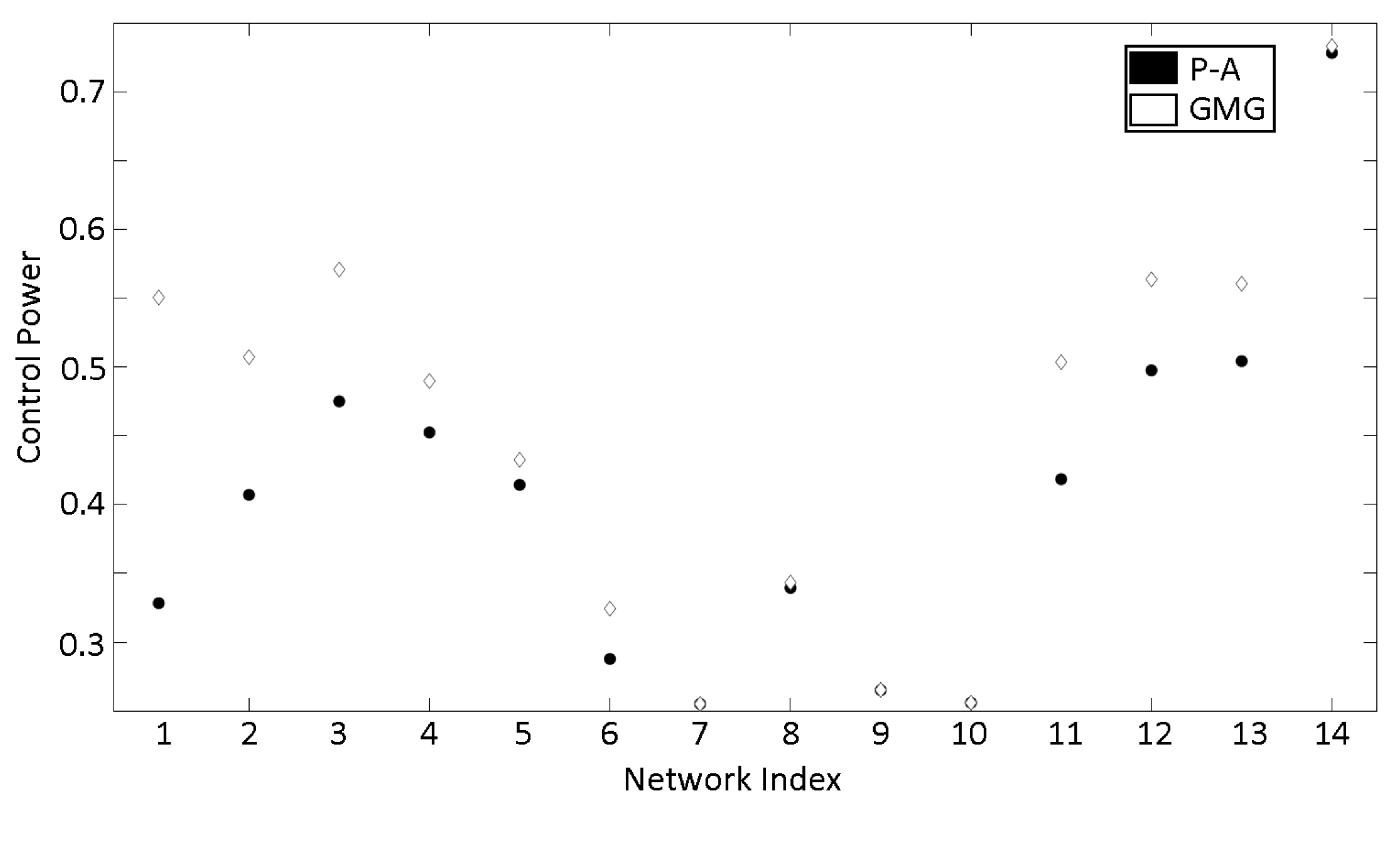}}
\caption{Control Power of networks following control strategy of the BA model and the GMG model with complete information of training set.}
\label{str1}
\end{figure}

\begin{figure}[htb]
\centering
\centerline{\includegraphics[width=8.5cm]{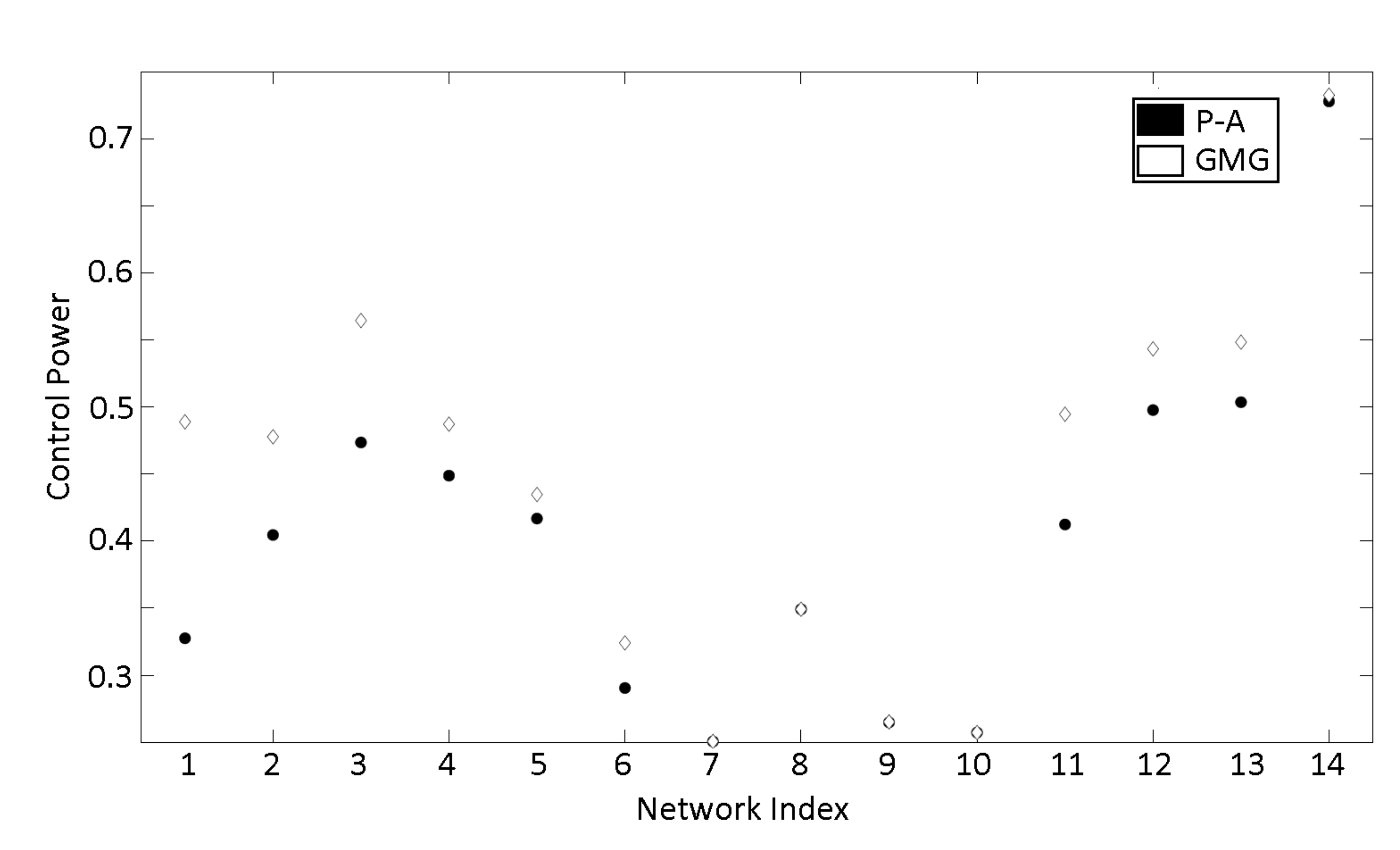}}
\caption{Control Power of networks following control strategy of the BA model and the GMG model with partial information of training set.}
\label{str2}
\end{figure}

From the experimental results, we can see that, with the same budget on the control set, the control strategy of the GMG model generates a higher control power than that of the BA model. Although the performance of the GMG model with partial information of the training set on control power estimation is significantly worse than the one with full information, their performance on control strategy are very close to each other.
\section{Conclusion and Future Research}
\label{cfr}
In this paper, we introduced an information flow model to simulate the information flow in a social network. Two social network models are used to optimize the control strategy, and compute the converged beliefs of agents in a social network. Compared to a direct calculation of the converged beliefs, these two models use less information and require less computational power, but still provide results with good accuracy. In addition, the GMG Model outperforms the BA model both in belief estimation accuracy and belief control strategy, since it has a more realistic assumption and uses more information.

Future work includes better machine learning techniques for computing $\alpha$, potentially theoretical calculation for $\alpha$ from network models, and more complicated information flow models. Other future open questions include temporal variability of social network features and their influence on belief estimation.

\appendices
\section{Network Synthesis in Barab\'{a}ási-Albert Model}
\label{pro1}
The input of the synthesis process in the BA model is the total number of nodes $N$ and the average number of edges attached to each new incoming node $m$. At the very beginning of the process, when the time step $t=1$, there are $m+1$ nodes in the network and they form a fully connected network. The first node is then introduced to the network with $m$ edges attached. Each edge will choose a node to connect in the network based on its degree and clustering coefficient. At time step $t$, when the ${\hat{m}}^{th}$ edge of the $m+1+t$ node is seeking  another node to attach to, an arbitrary node $i$ in the network is chosen with probability $^{1}p_i(t,\hat{m})$:
$$^{1}p_i(t,\hat{m}) = \frac{d_i(a+\gamma_i)^{\alpha}}{\Sigma_{j=1}^{m+t}d_i(a+\gamma_i)^{\alpha}},$$
according to Equation (\ref{PA}). After all $m$ edges of the first node are connected to the nodes in the network, the time step $t$ changes to $t = 2$, and the second node is introduced. At an arbitrary time step $t$, where $1<t\leq N-m$, a new node is introduced to the network with $m$ edges attached. Each edge will choose a node in the network with probability proportional to $^{1}p_i(t,\hat{m})$. The process ends when there are $N$ nodes in the network and, the edges brought by the last node are all connected with the nodes in the network. The degree distribution of networks generated by this process will follow a power law distribution\cite{C1}. The flow chart of this process is shown in Figure (\ref{FlowChartP}).
 
\begin{figure}[htb]
\centering
\centerline{\includegraphics[width=8.5cm]{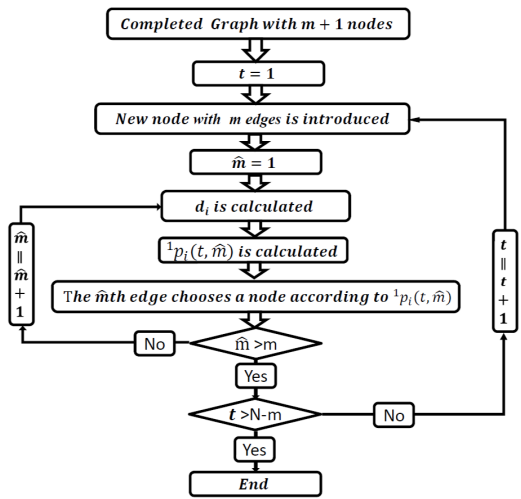}}
\caption{Flow chart for network synthesis in a BA model.}
\label{FlowChartP}
\end{figure}
}

\section{Network Synthesis in a Generalized Markov Graph Model}
\label{pro2}
In a GMG model, the network synthesis includes the following steps\cite{C12}. The input of the synthesis process is: the total number of nodes $N$, the average number of edges attached to each node $m$, and the clustering weight $\alpha$. At the very beginning of the process, at time step $t=1$, there are $m+1$ nodes in the network and they form a fully connected network. The first node is then introduced to the network with $m$ edges attached. Each edge will choose a node in the network to connect to by its degree and clustering coefficient. At time step $t$, when the ${\hat{m}}^{th}$ edges of the $m+1+t$ node is seeking another node to attach to, an arbitrary node $i$ in the network is chosen with probability $^{2}p_i(t,\hat{m})$:
$$^{2}p_i(t,\hat{m}) = \frac{d_i(a+\gamma_i)^{\alpha}}{\Sigma_{j=1}^{m+t}d_i(a+\gamma_i)^{\alpha}},$$
according to Equation (\ref{PAG}). After all $m$ edges of the first node are connected to the nodes in the network, the time step $t$ changes to $t = 2$ and the second node is introduced. At an arbitrary time step $t$, where $1<t\leq N-m$, a new node is introduced to the network with $m$ edges attached. The $\hat{m}^{th}$ edge will then choose a node in the network with probability $^{2}p_i(t,\hat{m})$. The process ends when there are $N$ nodes in the network and the edges brought by the last node are all connected to the nodes in the network. The degree distribution of networks generated by a GMG model can change according to the choice of value of $\alpha$\cite{C2}. The flow chart of this process is shown in Figure (\ref{FlowChartG}).
 
\begin{figure}[htb]
\centering
\centerline{\includegraphics[width=8.5cm]{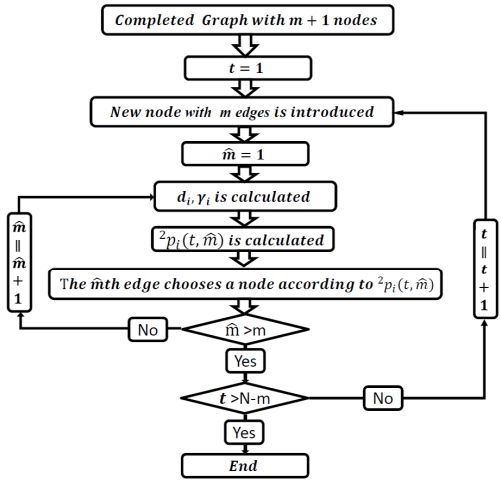}}
\caption{Flow chart for network synthesis in a GMG model.}
\label{FlowChartG}
\end{figure}

\section{Proof for Theorem ~\ref{AijPT}}
\label{App1}

The master equation and the solution of the master equation for the BA model is introduced in Barab\'{a}ási(2002)\cite{C1}. As a reminder, we restate it here. 

In a BA model, at each time step $t$, where $t$ lies in the range $[1,N]$, and $N$ is the highest node index at the last time step in the network, a new node with $m$ edges is added to the network. We denote this node by its joining time $t$. The probability of an existing node $i$ in the network be connected to the new node $t$ is proportional to its degree $d_i(t)$ at time $t$. This means an increasing rate of degree of an existing node $i$ in the network is proportional to its degree, as shown in the master equation, Equation (\ref{PdePA})\cite{C1}:

\begin{equation}
\label{PdePA}
\frac{\partial d_i(t)}{\partial t} = m \frac{d_i(t)}{\Sigma_{j=1}^{t}d_j(t)}.
\end{equation}

And the summation on the denominator of Equation (\ref{PdePA}) is known to be $2mt$ , as in each time step $m$ edges are added to the network. So we can rewrite Equation (\ref{PdePA}) as,
\begin{equation}
\label{PdePA1}
\frac{\partial d_i(t)}{\partial t} = \frac{d_i(t)}{2t},
\end{equation}
and the solution to this equation is\cite{C1}:
\begin{equation}
\label{PdePAS}
d_i(t) = m\sqrt{{t}/{i}}
\end{equation}

In the BA model, the probability of a node $i$ be connected with node $j$ is the probability of the following event happening: at time step $max(i,j)$, node $min(i,j)$ is chosen to be attached to one of the $m$ edges brought by the new node. According to the basic assumption of the model and Equation (\ref{PdePA}), such a probability will be:

\begin{equation}
\label{Sol1}
^{1}P_{i,j} = {d_{min(i,j)}(max(i,j))}/{(2 \space max(i,j))}.
\end{equation}

According to Equation (\ref{PdePAS}), the degree of node $min(i,j)$ at time step $max(i,j)$ is:

\begin{equation}
\label{minmax}
d_{min(i,j)} =  m\sqrt{{max(i,j)}/{min(i,j)}},
\end{equation}

Plugging  Equation (\ref{minmax}) into Equation (\ref{Sol1}), yields:
\begin{equation}
\label{Sol2}
^{1}P_{i,j} = {m}/{2} \sqrt{{1}/{(i j)}}.
\end{equation}

At an arbitrary time step $t$, the degree of node $i$ and $j$ are: $m\sqrt{{t}/{i}}$ and $m\sqrt{{t}/{j}}$. The degree summation at time step $t$ is $\Sigma_{k=1}^{t} d_k(t) = 2mt$, which together with Equation (\ref{Sol2}) leads to:

\begin{equation}
\label{Sol3}
^{1}P_{i,j} = {d_i(t) d_j(t)}/{\Sigma_{k=1}^{t} d_k(t)}.
\end{equation}

If the number of nodes in the network reaches N, which means the time step is at: $t = N$, and we omit the time notation for degree $d_i(t)$, we get:
 
\begin{equation}
\label{Sol3}
^{1}P_{i,j} = {d_i d_j}/{\Sigma_{k=1}^{N} d_k}.
\end{equation}
This concludes the proof. \hfill  $\blacksquare$

\section{Proof for Theorem ~\ref{Thm2}}
\label{App2}
According to Equation (\ref{CBCCS}) and the definition of control vector $\textbf{V}$, in a network $G$ of $N$ nodes with a control strategy, the expected belief of an arbitrary non-control node $i$, where $i\not\in \mathbf{C}$, at an arbitrary time step $T$, where $T\in\mathbb{Z}^{+}$, will be,

\begin{equation}
\label{thm21}
\begin{split}
\overline{^{1}B}_{i,t} = &\Sigma_{t=0}^{T-1} \Sigma_{j=1}^{c} B^*_{\theta_j} \overline{(A^{*}\times M)^{t}}_{{\theta_j},i}\\
 + &\Sigma_{t=0}^{T-1}\Sigma_{j=c+1}^{N} \overline{w^{*}}_{\theta_j}  \overline{(A^{*}\times M)^{t}}_{{\theta_j},i}.
\end{split}
\end{equation}

The matrix elements $(A^{*}\times M^{t})_{i,j}$, with $A^{*}_{i,j} = A_{i,j}/(1+d_j)$ and the definition of control matrix $\textbf{M}$, can be written as:

\begin{equation}
\label{thm211}
\begin{split}
&(A^{*}\times M)^{t}_{i,j}\\
= &\Sigma_{k_1=c+1}^{N}\dots \Sigma_{k_{t-1}=c+1}^{N} \frac{A_{i,{\theta_{k_1}}}}{1+d_{\theta_{k_1}}}\frac{A_{{\theta_{k_1}},{\theta_{k_2}}}}{1+d_{{\theta_{k_2}}}} \dots  \frac{A_{{\theta_{k_{t-1}}},j}}{1+d_{{\theta_{k_j}}}}.
\end{split}
\end{equation}
%

To calculate the expected value of $(A^{*}\times M^{t})_{i,j}$, we combine the expected value of $A_{i,j}$, which equates $P_{i,j}$ as shown in Equation (\ref{AijP}), together with Equation (\ref{thm211}) to yield



\begin{equation}
\label{thm213}
\overline{(A^{*}\times M)^{t}}_{i,j}
= \frac{d_i d_j}{(1+d_j)(\Sigma_{n=1}^{N}d_n)}
\left(\frac{\Sigma_{k=c+1}^{N}\frac{{d_{\theta_k}}^2}{1+d_{\theta_k}}}{\Sigma_{n=1}^{N}d_n}\right)^{t-1}.
\end{equation}

Let $$\beta_1 = {\Sigma_{k=c+1}^{N}\frac{{d_{\theta_k}}^2}{1+d_{\theta_k}}}/{\Sigma_{n=1}^{N}d_n}.$$ Because degree $d_k$ is a positive integer: $d_k \geq 1$, we have,

 $$d_k > \frac{d_k^2}{1+d_k},$$ and 
 $$\Sigma_{n=1}^{N}d_n > \Sigma_{k=c+1}^{N}{{d_{\theta_k}}^2}/{(1+d_{\theta_k})}.$$

So ${\beta_1}<1,$
and the larger $c$ of the control group is, the smaller ${\beta_1}$ is.  

Inserting Equation (\ref{thm213}) into Equation (\ref{thm21}), and seting $T=\infty$ yields

\begin{equation}
\label{thm214}
\begin{split}
\overline{^{1}B}_{i,\infty} = \Sigma_{j=1}^{c} {B^*_{\theta_j} d_i d_{\theta_j}}
\Sigma_{t=0}^{\infty}{\beta_1}^{t-1}/{(1+d_i)/(\Sigma_{n=1}^{N}d_n)}\\
 + \Sigma_{j=c+1}^{N} {{w_{\theta_j}}d_i d_{\theta_j} }
\Sigma_{t=0}^{\infty}{\beta_1}^{t-1}/({(1+d_i){(1+d_{\theta_j})}(\Sigma_{n=1}^{N}d_n)}).
\end{split}
\end{equation}

As ${\beta_1} < 1$, we can rewrite $\Sigma_{t=0}^{\infty}{\beta_1}^{t-1}$ as $\frac{1}{1-{\beta_1}}$ and Equation (\ref{thm214}) becomes

\begin{equation}
\overline{^{1}B}_{i,\infty} = \frac{1}{\Sigma_{n=1}^{N}  d_n}\frac{d_i}{1+d_i} \frac{{\Sigma_{j=1}^{c}B^*_{\theta_j} d_{\theta_j}} + {\Sigma_{j=c+1}^{N}\frac{\overline{{w}}_{\theta_j}}{1+d_{\theta_j}} d_{\theta_j}}}{1-{\beta_1}}.
\end{equation}

This concludes the proof. \hfill $\blacksquare$

\section{Proof for Theorem ~\ref{ThmG}}
\label{App3}
The proof of Theorem ~\ref{ThmG} will be similar to that of Theorem ~\ref{Thm2}. Based on Equation (\ref{CBCCS}) and the definition of control vector $\textbf{V}$, in a network $G$ of $N$ nodes with control strategy, the expected belief of an arbitrary non-control node $i$, where $i\not\in \mathbf{C}$, at an arbitrary time step $T$, where $T\in\mathbb{Z}^{+}$, will be:

\begin{equation}
\label{thm31}
\begin{split}
\overline{^{2}B}_{i,t} = &\Sigma_{t=0}^{T-1} \Sigma_{j=1}^{c} B^*_{\theta_j} \overline{(A^{*}\times M)^{t}}_{{\theta_j},i}\\
 + &\Sigma_{t=0}^{T-1}\Sigma_{j=c+1}^{N} \overline{w^{*}}_{\theta_j}  \overline{(A^{*}\times M)^{t}}_{{\theta_j},i}.
\end{split}
\end{equation}

The matrix element $(A^{*}\times M^{t})_{i,j}$, with $A^{*}_{i,j} = A_{i,j}/(1+d_j)$ and the definition of control matrix $\textbf{M}$, may be written as:

\begin{equation}
\label{thm311}
\begin{split}
&(A^{*}\times M)^{t}_{i,j}\\
= &\Sigma_{k_1=c+1}^{N}\Sigma_{k_2=c+1}^{N} \dots \Sigma_{k_{t-1}=c+1}^{N} \frac{A_{i,k_1}}{1+d_{k_1}}\frac{A_{k_1,k_2}}{1+d_{k_2}} \dots \frac{A_{k_{t-1},j}}{1+d_{k_j}}.
\end{split}
\end{equation}

$\overline{(A^{*}\times M)^{t}}_{i,j}$ is then calculated from the expected value of $A_{i,j}$, which equates $^{2}P_{i,j}$ as shown in Equation (\ref{APij}), and Equation (\ref{thm311}):



\begin{equation}
\label{thm313}
\begin{split}
&\overline{(A^{*}\times M)^{t}}_{i,j}\\
= \ &\frac{d_i(1+\gamma_i)^{\alpha} d_{j}(1+\gamma_{j})^{\alpha}(\Sigma_{n=1}^{N}d_n)}{(1+d_j)\eta}\\
&\left({\Sigma_{k=c+1}^{N}\frac{{(d_{\theta_{k}}(1+\gamma_{\theta_{k}})^{\alpha})}^2}{(1+d_{\theta_k})\eta}}{\Sigma_{n=1}^{N}d_n}\right)^{t-1}.
\end{split}
\end{equation}

Let $$\beta_2 = \frac{\Sigma_{k=c+1}^{N}\frac{{(d_{\theta_{k}}(1+\gamma_{\theta_{k}})^{\alpha})}^2}{1+d_{\theta_k}}}{\eta}{\Sigma_{n=1}^{N}d_n}.$$

Again inserting Equation (\ref{thm313}) into Equation (\ref{thm31}), and setting $T=\infty$ yields

\begin{equation}
\label{thm314}
\begin{split}
&\overline{^{2}B}_{i,\infty}\\
= &\  \Sigma_{j=1}^{c} \frac{B^*_{\theta_j} d_i(1+\gamma_i)^{\alpha} d_{{\theta_j}}(1+\gamma_{{\theta_j}})^{\alpha}(\Sigma_{n=1}^{N}d_n)}{(1+d_i)\eta}\Sigma_{t=0}^{\infty}{\beta_1}^{t-1}\\
& + \Sigma_{j=1}^{c} \frac{\frac{\overline{w}_{\theta_j}}{1+d_{\theta_j}} d_i(1+\gamma_i)^{\alpha} d_{{\theta_j}}(1+\gamma_{{\theta_j}})^{\alpha}(\Sigma_{n=1}^{N}d_n)}{(1+d_i)\eta}\Sigma_{t=0}^{\infty}{\beta_1}^{t-1}.
\end{split}
\end{equation}

If ${\beta_2} < 1$, we can rewrite $\Sigma_{t=0}^{\infty}{\beta_2}^{t-1}$ as $\frac{1}{1-{\beta_2}},$ and Equation (\ref{thm314}) becomes

\begin{equation}
\begin{split}
\overline{^{2}B}_{i,\infty} = &\frac{\Sigma_{n=1}^{N}  d_n}{\eta}\frac{d_i(1+\gamma_i)^{\alpha}}{1+d_i}\\
&\frac{{\Sigma_{j=1}^{c} B^*_{\theta_j} d_{{\theta_j}}(1+\gamma_{{\theta_j}})^{\alpha}} + {\Sigma_{j=c+1}^{N}\frac{\overline{{w}}_{\theta_j}}{1+d_{\theta_j}} d_{{\theta_j}}(1+\gamma_{{\theta_j}})^{\alpha}}}{1-{\beta_2}}
\end{split}
\end{equation}

This concludes the proof. \hfill $\blacksquare$

\section{Proof for Theorem ~\ref{ConT1}}
\label{App4}

Consider an arbitrary control set $\mathbf{C^{'}} = \{ \theta^{'}_1, \theta^{'}_2, \dots, \theta^{'}_c\}$  with corresponding degrees $\{d_{\theta^{'}_1}, d_{\theta^{'}_2}, \dots, d_{\theta^{'}_c}\}$. Without loss of generality,  we set $d_{\theta^{'}_1} \geq d_{\theta^{'}_2} \geq \dots \geq d_{\theta^{'}_c}$. For control set $\mathbf{C_o} = \{ {\theta_o}_1,{\theta_o}_2, \dots, {\theta_o}_c \}$, the corresponding degrees satisfy $d_{{\theta_o}_i} \geq d_{{\theta_o}_j}$ if $i\leq j$, $1\leq i,j \leq N$, so, for $1 \leq i\leq c$, we have

\begin{equation}
\label{eqdeg1}
d_{{\theta_o}_i} \geq d_{\theta^{'}_i}.
\end{equation}

Since $\overline{w}_i = 0$, for $i = 1, \dots, N$, $B^{*}_j = 1$, for $1\leq j \leq c$, according to Equation~(\ref{cpPAeq}), $^1 cp$ can be rewritten as:

\begin{equation}
^1 cp =  \Sigma_{i=1}^{N}\frac{d_i}{1+d_i} ({\Sigma_{j = 1}^{c}d_{\theta^{'}_j}})/({\Sigma_{k = 1}^{N}d_k - \Sigma_{m = c+1}^{N}\frac{{d_{\theta^{'}_m}}^2}{1 + d_{\theta^{'}_m}}}).
\end{equation}
Since $\Sigma_{i=1}^{N}\frac{d_i}{1+d_i}$ and $\Sigma_{k = 1}^{N}d_k$ are positive constants, the derivative of control power $^1 cp$ with respect to the degree of a controlled node $d_{\theta^{'}_i}$ is:

\begin{equation}
\frac{\partial ^1 cp}{\partial d_{\theta^{'}_i}} = D_1 \frac{\Sigma_{k = c+1}^{N}\frac{{d_{\theta^{'}_k}}}{1 + d_{\theta^{'}_k}} + \frac{\Sigma_{j=1}^{c}d_{\theta^{'}_j}}{(1+d_{\theta^{'}_i)^2}}}{D_2^2},
\end{equation}
where $D_1 = \Sigma_{j=1}^{N}\frac{d_j}{1+d_j}$, $D_2 = \Sigma_{k=1}^{N}\frac{d_k}{1+d_k} + \Sigma_{i = c+1}^{N}\frac{{d_{\theta^{'}_i}}}{1 + d_{\theta^{'}_i}}$.
As $D_1 > 0$, one can get:
\begin{equation}
\label{eqfr1}
\frac{\partial ^1 cp}{\partial d_{\theta^{'}_i}} > 0.
\end{equation}
From Equations~(\ref{eqdeg1}}) and (\ref{eqfr1}), one can get:
\begin{equation}
^1cp(\mathbf{C_o}) \geq {^1cp(\mathbf{C^{'}})}.
\end{equation}

This concludes the proof. \hfill $\blacksquare$

\section{Proof for Theorem ~\ref{ConT2}}
\label{App5}

Define the weight $\xi_i$ of node $i$ as: $\xi_i = d_i(1+\gamma_i)^\alpha$. 
Consider an arbitrary control set $\mathbf{C^{'}} = \{ \theta^{'}_1, \theta^{'}_2, \dots, \theta^{'}_c\}$  with corresponding weight $\{d_{\xi^{'}_1}, d_{\xi^{'}_2}, \dots, d_{\xi^{'}_c}\}$. Without loss of generality, we set $d_{\xi^{'}_1} \geq d_{\xi^{'}_2} \dots \geq d_{\xi^{'}_c}$. For control set $\mathbf{^2 C_o} = \{ {\theta^{\dagger}_o}_1,{\theta^{\dagger}_o}_2, \dots, {\theta^{\dagger}_o}_c \}$, the corresponding weights satisfy $\xi_{{\theta^{\dagger}_o}_i} \geq \xi_{{\theta^{\dagger}_o}_j}$ if $i\leq j$, $1\leq i,j \leq N$, so, for $1 \leq i\leq c$, we have
\begin{equation}
\label{eqdeg2}
\xi_{{\theta^{\dagger}_o}_i} \geq \xi_{\theta^{'}_i}.
\end{equation}

Since $\overline{w}_i = 0$, for $i = 1, \dots, N$, and $B^{*}_j = 1$, for $1\leq j \leq c$, according to Equation~(\ref{cpGeq}) $^2 cp$ can be rewritten as:

\begin{equation}
^2 cp = K_1\frac{\Sigma_{j=1}^{c}\xi_{\theta^{'}_j}}{K_2 + K_3 \Sigma_{k=1}^{c}\frac{{\xi_{\theta^{'}_k}}^2}{1+d_{\xi_{\theta^{'}_k}}}},
\end{equation}

where $K_1 = \frac{1}{N} \Sigma_{i=1}^{N}\frac{\xi_i}{1+d_i} \Sigma_{j=1}^{N}d_j$, 

$K_2 = \Sigma_{i=1}^{N}{\xi_i}^2 - \Sigma_{j=1}^{N}{\xi_j} - \Sigma_{k=1}^{N}{d_k}\Sigma_{m=1}^{N}\frac{{\xi_m}^2}{1+d_m}$, 

$K_3 = \Sigma_{j=1}^{N}{d_j}. $

The derivative of control power $^2 cp$ with respect to the degree of a controlled node $d_{\theta^{'}_i}$ is:

\begin{equation}
\begin{split}
\frac{\partial ^2 cp}{\partial d_{\theta^{'}_i}} = \frac{K_4}{{K_5}^2} [ {(\frac{1}{\beta_2}-1)\Sigma_{j=c+1}^{N}\frac{{\xi_{\theta^{'}_j}}^2}{1+d_{\theta^{'}_j}} - \Sigma_{k=1}^{c}\xi_{\theta^{'}_k}(1+\gamma_{\theta^{'}_i})^{\alpha}} \\+ \Sigma_{m=1}^{c}\xi_{\theta^{'}_m}\frac{(1+\gamma_{\theta^{'}_i})^{\alpha}}{(1+d_{\theta^{'}_i})^2} ]
\end{split}
\end{equation}

where $K_4 = \frac{1}{N}\Sigma_{i=1}^{N}\frac{\xi_i}{1+d_i}$,

$K_5 = \frac{\Sigma_{i = 1}^{N}{\xi_i}^2 - \Sigma_{j = 1}^{N}{\xi_j}}{K_3} - \Sigma_{k=1}^{N}\frac{{\xi_k}^2}{1+d_k}$.

Since $K_4 > 0$, we will have
\begin{equation}
\frac{\partial ^2 cp}{\partial d_{\theta^{'}_i}} > 0,
\end{equation}
assuming that $\beta_2$ satisfies:
\begin{equation}
\label{beta2con}
\frac{1}{\beta_2} > 1 + max(1,2^{\alpha})\frac{\Sigma_{j=1}^{c} d_{{\theta^{\dagger}_o}_j}(1+\gamma_{{\theta^{\dagger}_o}_j})^{\alpha}}{\Sigma_{k = c+1}^{N} \frac{(d_{{\theta^{\dagger}_o}_k}(1+\gamma_{{\theta^{\dagger}_o}_k})^{\alpha})^2}{1+d_{{\theta^{\dagger}_o}_k}}}.
\end{equation}

Considering Equation~(\ref{eqdeg2}}), one can get:
\begin{equation}
^2cp(\mathbf{C^{\dagger}_o}) \geq {^2 cp(\mathbf{C^{'}})},
\end{equation}
if Equation~(\ref{beta2con}) holds.

This concludes the proof. \hfill $\blacksquare$



\bibliographystyle{plain}
\bibliography{IP}
\begin{IEEEbiography}[{\includegraphics[width=1in,height=1.25in,clip,keepaspectratio]{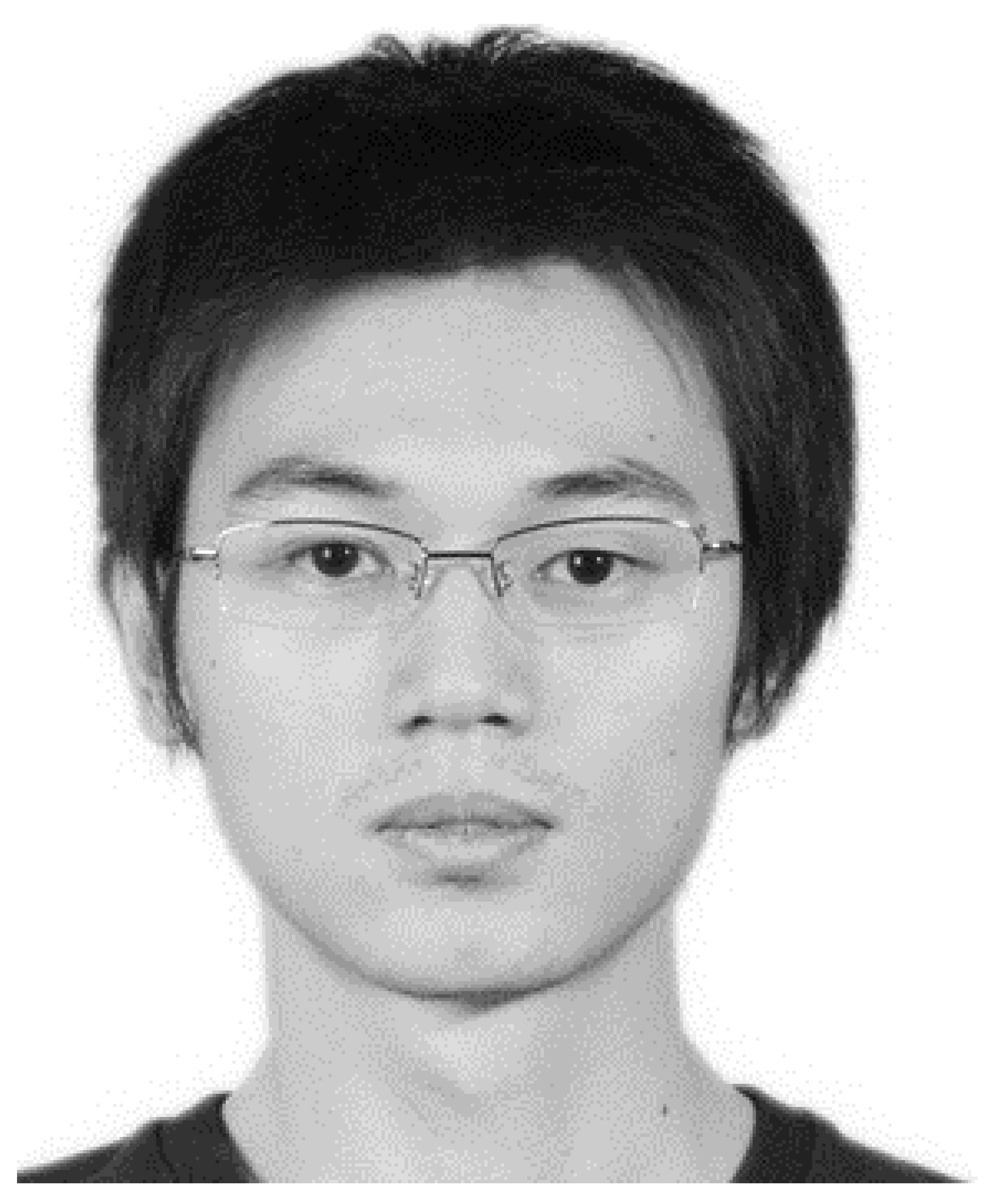}}]{Tian Wang}
received his
B.S. (2008) degree in Physics
from the University of Science and Technology of China and the M.Sc. (2011) degree in electrical engineering and Physics from North Carolina State University, where he been a Ph.D. student in Physics since 2008. His main research interest is in modelling and analysis of social network and information flow on social network.
\end{IEEEbiography}


\begin{IEEEbiography}[{\includegraphics[width=1in,height=1.25in,clip,keepaspectratio]{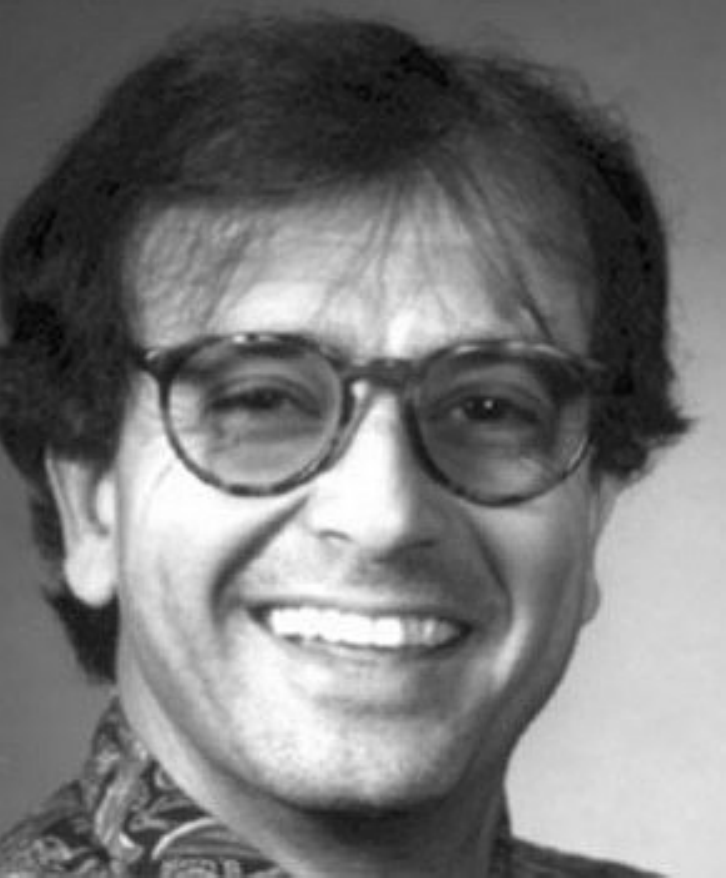}}]{Hamid Krim}
received his degrees in Electrical Engineering. As a member of technical staff at AT\&T Bell Labs, he has worked in the area of telephony and digital communication systems/subsystems. In 1991 he became
a NSF Post-doctoral scholar at Foreign Centers of Excellence (LSS Sup-elec/Univ. of Orsay, Paris, France). He subsequently joined the Laboratory
for Information and Decision Systems, MIT, Cambridge, MA, as a Research
Scientist performing/supervising research in his area of interest, and later as a faculty in the ECE dept. at North Car. State Univ. in Raleigh, N.C. in
1998. He is an original contributor and now an affiliate of the Center for
Imaging Science sponsored by the Army. He also is a recipient of the NSF
Career Young Investigator award.
He is on the editorial board of the IEEE Trans. on SP and regularly
contributes to the society in a variety of ways. His research interests are in
statistical estimation and detection and mathematical modeling with a keen
emphasis on applications.
\end{IEEEbiography}

\begin{IEEEbiography}[{\includegraphics[width=3.0in,height=1.3in,clip,keepaspectratio]{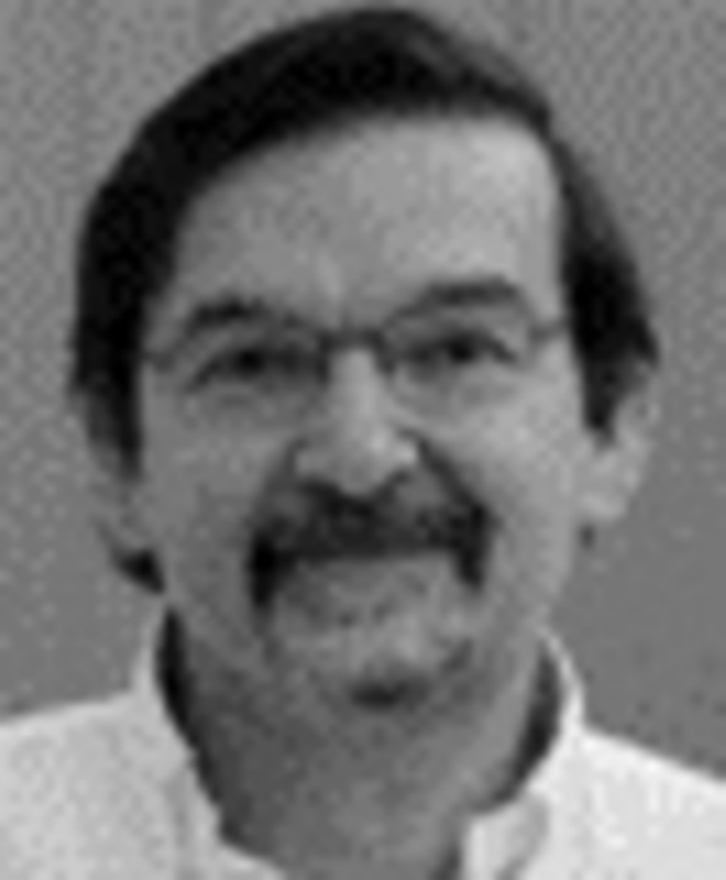}}]{Yannis Viniotis}
received his Ph.D. from the University of Maryland, College Park, 
in 1988 and is currently Professor, Department of Electrical and Computer Engineering 
at North Carolina State University. 
Dr. Viniotis is the author of over one hundred technical publications, 
including two engineering textbooks. 
He has served as the cochair of two international conferences in computer networking. 
His research interests include Service-Oriented Architectures, service engineering, 
and design and analysis 
of stochastic algorithms. Dr. Viniotis was the cofounder of Orologic, 
a successful startup networking company in Research Triangle Park, NC, 
that specialized in ASIC implementation of integrated traffic management 
solutions for high-speed networks.
\end{IEEEbiography}

\end{document}